\documentclass[useAMS,usenatbib]{mn2e}
\usepackage{epsfig}
\usepackage{eufrak}
\usepackage{amsmath}
\usepackage{rotating}
\usepackage{graphicx}
\usepackage{color}

\title[Testing PDR models against ISO fine structure line data for extragalactic sources.]
{Testing PDR models against ISO fine structure line data for extragalactic sources.}
\author[M. ~Vasta et al.]
{M. Vasta$^1$, M. J. Barlow$^1$, S. Viti$^1$, J. A. Yates$^1$, T. A. Bell$^{2}$.\\
$^1$Department of Physics and Astronomy, University College London,
      Gower Street, London WC1E 6BT, UK\\
$^2$Caltech, Department of Physics, MC 320- 47, Pasadena, CA 91125, USA}
\date{Accepted ???, Received ???; in original form \today}

\begin{document}
\maketitle

\begin{abstract}

Far-infrared [C~{\sc ii}] 158-$\mu$m, [O~{\sc i}] 145-$\mu$m and [O~{\sc i}] 63-$\mu$m fine structure emission line fluxes were measured from archival Infrared Space Observatory Long Wavelength Spectrometer spectra of 46 extragalactic sources, with 28 sources providing detections in all three lines. For 12 of the sources, the contribution to  the [C~{\sc ii}] 158-$\mu$m line flux from H~{\sc ii} regions could be estimated from their detected [N~{\sc ii}] 122-$\mu$m line fluxes. The measured [C~{\sc ii}]/[O~{\sc i}] and [O~{\sc i}] 63/145-$\mu$m line flux ratios were compared with those from a grid of PDR models previously computed using the UCL\_PDR code. Persistent offsets between the observed and modelled line ratios could be partly attributed to the effects of [O~{\sc i}] 63-$\mu$m self-absorption. Using the SMMOL code, we calculated  model [O~{\sc i}] line
profiles and found that the strength of the [O~{\sc i}] 63-$\mu$m line was reduced by 20-80\%, depending on the PDR parameters. We conclude that high PDR densities and radiation field strengths, coupled with the effects of [O~{\sc i}] 63-$\mu$m self-absorption, are likely to provide the best match to the observed line flux ratios.

\end{abstract}

\section{Introduction}

Studies of our own Galaxy and observations of external galaxies have 
suggested that stellar ultraviolet radiation can ionize vast volumes of a 
galaxy and that far-ultraviolet radiation impinging on neutral cloud 
surfaces is responsible for a large fraction of the observed far-infrared 
(FIR) spectral line emission that cools the gas \citep{Crawford85}. The 
relative contributions from different gas phases can be estimated from 
observations of several FIR forbidden lines. Fine structure (FS) emission 
lines can be used as tracers of nebular conditions such as density, 
excitation and ionization. By virtue of their different excitation 
potentials and critical densities, FS emission lines provide an insight 
into the energetics and chemical composition of the regions from which 
they originate. The FIR fine structure lines of abundant species, such as 
[C {\sc ii}] and [O {\sc i}], have long been recognised as one of the most 
important coolants in the interstellar medium (ISM).  In particular, the 
emission from singly ionized carbon [C {\sc ii}], at $\lambda$=158 $\mu$m, 
is known to trace warm and dense photodissociation regions (PDRs) and is 
used to trace the gas surrounding active star-forming regions. Carbon is 
the fourth most abundant element and has a lower ionization potential 
(11.26 eV) than hydrogen. For this reason it is predominantly in the form 
of C$^{+}$ in the surface layers of PDRs, where photoionization of neutral 
carbon is effective, but can potentially also be excited in ionized 
regions. [C {\sc ii}] 158$\mu$m emission is reasonably easy to excite, 
with a transition energy of 92 K. The line is widely observed and is 
usually optically thin. The depth of the C$^{+}$ zones in neutral clouds 
is generally determined by dust extinction of carbon ionizing 
photons and often extends to A$_{v}$ $\geq $4, although C$^{+}$ decreases 
in abundance for A$_{v}$$\geq $1 \citep{Bell:06}. 

Another important species 
is oxygen, which has an ionization potential of 13.62 eV, slightly above 
that of hydrogen. This means that the ionization structure of 
O$^{+}\backslash$O$^{0}$ closely follows the ionization structure of 
hydrogen. Also, the charge exchange of neutral oxygen in two-body 
recombination with hydrogen cannot be neglected as it affects the 
ionization balance. The incident FUV radiation maintains a significant 
abundance of atomic oxygen deep into the cloud through photodestruction of 
CO and O$_{2}$. All oxygen not incorporated into CO can remain in atomic 
form to depths as large as A$_{v}$=10 mag under strong FUV irradiation 
\citep{Sternberg95}.The fraction of oxygen at magnitudes greater 
than 10 depends on the physical conditions in the cloud (i.e. its density, 
metallicities and radiation field strength), but essentially if the gas 
density is $>$10$^{5}$ cm$^{-3}$ at these depths, the oxygen not locked 
in CO is depleted on the grains. However for the purpose of these 
calculations, the code that we use, UCL\_PDR, does not take into account 
freeze out 
reactions. Atomic oxygen has two fine structure transitions, at 63 and 
145 $\mu$m. The [O {\sc i}] line at 63$\mu$m has an excitation energy 
corresponding to 228 K, is emitted predominantly by warm and dense gas, although it can also have an ionized gas component 
\citep{Aannestad03}, 
and can act as a coolant in neutral hydrogen regions. It may 
also be a heating agent as O {\sc i} can heat the gas through the 
collisional de-excitation of the dust continuum radiation-excited $^3P_1$ 
fine structure level \citep{TH85}. However this mechanism is not well 
studied. This line is optically thick in some cases \citep{Abel07} and 
its 
optical depth can be estimated by comparing the [O {\sc i}] line flux at 
63$\mu$m to that of the [O {\sc i}] line at 145$\mu$m \citep{Abel07}. The 
[O {\sc 
i}] 145$\mu$m line can be harder to detect: in fact the relative faintness 
of this line diminishes its capabilities as a robust PDR diagnostic. The 
lower level of the 145$\mu$m line is not the ground state of O$^{0}$, 
meaning that [O {\sc i}] 145$\mu$m emission is usually optically thin 
\citep{Abel07}. The 145$\mu$m line has an excitation energy of 326 K, 
almost 100 K higher than that of the 63$\mu$m line, so the ratio of the 
two line intensities is sensitive to the gas temperature for T$\leq$300 K. 
Since the strength of the FUV radiation field governs the heating of the 
gas, this line ratio is also a diagnostic of the incident FUV flux 
\citep{TH85}.
 
In the presence of foreground cold and tenuous gas the [O {\sc i}] 
63{$\mu$}m line can show both emission and absorption components 
\citep{TH85,Liseau06,Gonzo04}. Nevertheless observations of the FIR 
fine structure emission lines of [O {\sc i}] and [C {\sc ii}] can be used 
as diagnostics to infer the physical conditions in the emitting gas, such 
as temperatures, densities and radiation field strengths, by comparing 
with models of photodissociation regions \citep{TH85} and H{\sc ii} 
regions \citep{Rubin91}. In the past, several models have been used to 
interpret infrared, submillimeter and millimeter line observations of 
neutral gas in our own Galaxy as well as in the central regions of nearby 
starburst galaxies \citep[e.g.][and references 
therein]{Mao00,Seaquist00,Wild92}.
 
In this paper we will use FIR line fluxes measured from the ISO archive 
for a sample of extragalactic sources. We present the fluxes of fine 
structure emission lines observed by ISO from 46 sources and we estimate 
the bulk properties of the gas in these external galaxies using the {\sc 
UCL\_PDR} model of \citet{Bell05} and the radiative transfer code SMMOL 
\citep{Rawlings01}.

The paper is structured as follows: in Section 2 we give details of the 
sources selected and the fine structure emission line fluxes we measured. 
In Section 3 we describe the {\sc UCL\_PDR} models. In Section 4 we 
evaluate the sensitivity of the {\sc UCL\_PDR} models to variations in the 
physical parameters. In Section 5 we will consider the contribution of H 
{\sc ii} regions to fine structure emission lines. The results are given 
in Section 6, and there we discuss their interpretation in the context 
of our SMMOL line profile calculations. Concluding 
remarks are given in Section 7.

\section{The ISO data}
\label{sec:data}
All the FIR data included in this paper were acquired using the ISO Long 
Wavelength Spectrometer \citep[LWS;] []{Clegg96}, which had an estimated 
FWHM beam size of $\sim$ 80 arcsec.

We collected [C {\sc ii}]158{$\mu$}m, [O {\sc i}]145{$\mu$}m and 
63{$\mu$}m emission line fluxes from ISO LWS archival observations for 46 
external galaxies. The [C {\sc ii}]158{$\mu$}m, [O {\sc i}]145{$\mu$}m, 
63{$\mu$}m emission line fluxes were measured from the spectra using the 
interactive package SMART (Spectroscopy Modeling Analysis and reduction 
Tool; \citet{Higdon04}.)

\begin{table*}
\begin{minipage}{180mm}
\scriptsize{
\caption{Line flux measurements, in units of 10$^{-14}$~W~m$^{-2}$, for 
extragalactic sources with ISO-LWS detections of all 3 FIR fine structure 
emission lines of [C~{\sc ii}] and [O~{\sc i}]. For each source, our own 
line flux measurements are listed in the first row, while literature 
values, if any, are listed in the second row 
\citep{Contursi02,Bergvall00,Colbert99,Fischer96,Luhm03,Malhotra01,Negishi01,Unger00,Comp08}. 
(n.a: not available)}
\label{sources}
\begin{tabular}{llllllllll}
\hline
{\bf Index}& {\bf Source}&{\bf Radial Vel}&{\bf TDT No. of }&{\bf [CII] 158$\mu$m}&{\bf [OI] 63$\mu$m}&{\bf [OI] 145$\mu$m}& {\bf \underline{[CII]$_{158}$}} & {\bf \underline{[OI]$_{63}$}}\\
& & {\bf (km s$^{-1}$)} &{\bf Observation }& & & &{\bf [OI]$_{63}$} &{\bf [OI]$_{145}$}\\
\hline
1 & {\bf IC 10} &  -348&  45700609&  0.966$\pm$0.034 & 0.591$\pm$0.025 & 0.0217$\pm$0.0032 & 1.64$\pm$0.13 & 27.23$\pm$5.17 \\
&&&&  0.763$\pm$0.021 &  0.65$\pm$0.02  &  0.024$\pm$0.004 &   1.17$\pm$0.07 &   27.08$\pm$5.34 \\
2 & {\bf Haro 11} &  6175&  54900720&  0.0410$\pm$0.0022 & 0.0954$\pm$0.0041 & $<$0.0086 & 0.43$\pm$0.04 & n.a \\
&&&&  0.039$\pm$0.001 &  0.092$\pm$0.018 &  0.003$\pm$0.001 & 0.42$\pm$0.09 & 30.6$\pm$16.2\\
3 & {\bf NGC 253} &243&  56901708&  4.618$\pm$0.126 &  3.47$\pm$0.16 & 0.461$\pm$ 0.074 & 1.33$\pm$0.09 & 7.53$\pm$1.55 \\
&&&&  5.19$\pm$1.04 &    3.76$\pm$0.75 &    0.52$\pm$0.11 & 1.38$\pm$0.55 & 7.23$\pm$2.97\\
4 & {\bf NGC 520} &2281&  77702295&  0.2188$\pm$0.0045 & 0.29$\pm$0.015 & 0.0133$\pm$0.0028 & 0.75$\pm$0.05 & 21.80$\pm$5.72 \\
&&&&  0.28$\pm$0.06 &    0.21$\pm$0.04 &  0.02$\pm$0.01 & 1.33$\pm$0.54 & 10.5$\pm$7.3\\
5 & {\bf Maffei 2} & -17&  85800682&  1.072$\pm$0.051 & 0.512$\pm$0.051 & 0.071$\pm$0.021 & 2.09$\pm$0.31 & 7.21$\pm$2.85 \\
&&&&  1.21$\pm$0.24 &    0.53$\pm$0.11 &    0.05$\pm$0.01 & 2.28$\pm$0.93& 10.6$\pm$4.3 \\
6 & {\bf NGC 1068} &1137&  60500401&  1.881$\pm$0.062 & 1.416$\pm$0.043 & 0.1522$\pm$0.0024 & 1.33$\pm$0.08 & 9.30$\pm$0.43 \\
&&&&  2.13$\pm$0.43 &    1.60$\pm$0.32 &    0.09$\pm$0.02 & 1.33$\pm$0.54 & 17.8$\pm$7.5 \\
7 & {\bf IC 342} &31&  64600302&  0.3146$\pm$0.0097 &  $<$0.117 & 0.0202$\pm$0.0049 & n.a & n.a\\
&&&&  n.a  &      n.a   &     n.a &     n.a &     n.a \\
8 & {\bf NGC 1482} &1916&  79600984&  0.571$\pm$0.022 & 0.362$\pm$0.025 &  $<$0.057 & 1.58$\pm$0.17& n.a\\
&&&&  0.655$\pm$0.013 & 0.318$\pm$0.063 & n.a &  2.06$\pm$0.45  &n.a   \\
9 & {\bf NGC 1569} &-104&  64600489&  0.652$\pm$0.027 & 0.657$\pm$0.025 & 0.0085$\pm$0.0019 & 0.99$\pm$0.08 & 77.3$\pm$20.2 \\
&&&&  0.674$\pm$0.134 &    0.589$\pm$0.119 &     n.a  & 1.14$\pm$0.46 & n.a \\
10 & {\bf NGC 1614} &4778&  85501010& 0.1953$\pm$0.0067 & 0.346$\pm$0.016 & 0.0179$\pm$0.0026& 0.56$\pm$0.05 & 19.33$\pm$3.70 &\\
&&&&  0.226$\pm$0.010   &  0.343$\pm$0.016    &       n.a   &     0.65$\pm$0.05  &     n.a\\
11 & {\bf NGC 2146} &893&  67900165&  2.479$\pm$0.078 & 1.756$\pm$0.091 & 0.164$\pm$0.023 & 1.41$\pm$0.12 & 10.70$\pm$2.06 &\\
&&&&  2.62$\pm$0.52 &    1.73$\pm$0.35 &    0.10$\pm$0.02 & 1.51$\pm$0.61 & 17.3$\pm$6.7\\
12 & {\bf NGC 2388} &4134&  71802360&  0.1473$\pm$0.0073 & 0.0954$\pm$0.0058 & 0.0032$\pm$0.0010  & 1.54$\pm$0.17& 29.81$\pm$11.13\\
&&&&  0.191$\pm$0.038 & 0.097$\pm$0.019 &     n.a  & 1.97$\pm$0.77 & n.a  \\
13 & {\bf M 82} &203&  65800611&  13.02$\pm$0.30 & 16.70$\pm$0.59 & 1.366$\pm$0.056 & 0.78$\pm$0.05 &12.23$\pm$0.93 &\\
&&&&  12.79$\pm$2.59 &   16.94$\pm$3.38 &    1.46$\pm$0.29 & 0.75$\pm$0.31 & 11.60$\pm$4.62 \\
14 & {\bf NGC 3256} &2804&  25200456&  1.124$\pm$0.044 & 1.178$\pm$0.033 &  $<$0.032 & 0.95$\pm$0.03 & n.a\\
&&&&  1.37$\pm$0.27 &    1.28$\pm$0.26 &     n.a & 1.07$\pm$0.43  & n.a \\
15 & {\bf IRAS 10565} &12921&  20200453&  0.0551$\pm$0.0076 & 0.0848$\pm$0.0069 &  $<$0.011 & 0.65$\pm$0.14 & n.a\\
&{\bf+2448}&&&&&&&\\
&&&&  0.047$\pm$0.009  &  0.076$\pm$0.008     &    n.a  & 0.61$\pm$0.18 & n.a \\
16 & {\bf NGC 3620} &1680&  27600981&  0.173$\pm$0.012 & 0.152$\pm$0.093 & $<$ 0.046 & 1.14$\pm$0.77& n.a\\
&&&&  0.249$\pm$0.049 &  0.164$\pm$0.029 &  0.029$\pm$0.006 & 1.52$\pm$0.57 & 5.65$\pm$2.17 \\
17 & {\bf NGC 3690} &3121&  18000704&  0.755$\pm$0.028 & 0.895$\pm$0.084 & 0.074$\pm$0.012 & 0.84$\pm$0.11 &12.09$\pm$3.10 \\
&&&&  0.86$\pm$0.17 &    0.83$\pm$0.17 &    0.05$\pm$0.005 & 1.03$\pm$0.41 & 16.6$\pm$5.1\\
18 & {\bf NGC 4039/9} &1641&  25301107&  0.378$\pm$0.011 & 0.412$\pm$0.079 & 0.0211$\pm$0.0041 & 0.92$\pm$0.20 & 19.53$\pm$7.54 \\
&&&&  0.37$\pm$0.01 &    0.34$\pm$0.07 &     n.a   & 1.09$\pm$0.25 & n.a    \\
19 & {\bf NGC 4102} &846&  19500584&  0.322$\pm$0.012 & 0.341$\pm$0.022 & 0.0141$\pm$0.0041 & 0.944$\pm$0.096 & 24.18$\pm$8.59\\
&&&&  0.289$\pm$0.049 &   0.269$\pm$0.049 &  0.022$\pm$0.004 & 1.07$\pm$0.38 & 12.22$\pm$4.45 \\
20 & {\bf NGC 4151} &995&  35800185&  0.0551$\pm$0.0043 & 0.401$\pm$0.032 & 0.0331$\pm$0.0047 & 0.14$\pm$0.02 & 12.11$\pm$2.71 \\
&&& 35300163&  0.074$\pm$0.006   &     0.376$\pm$0.029    &    n.a  &     0.19$\pm$0.031 &     n.a       \\
21 & {\bf NGC 4194} &2501&  19401369&  0.2094$\pm$0.0073 & 0.267$\pm$0.010 &  $<$0.0057 & 0.78$\pm$0.06 & n.a\\
&&&&  0.217$\pm$0.008     &    0.28$\pm$0.02      &   0.013$\pm$0.003  &  0.77$\pm$0.08 &     21.53$\pm$6.50    \\
22 & {\bf NGC 4449} &207&  23400120&  0.2383$\pm$0.0086 & 0.151$\pm$0.046 & 0.0187$\pm$0.0044 & 1.58$\pm$0.53 & 8.07$\pm$4.36\\
&&&&  0.278$\pm$0.008     &       0.133$\pm$0.013      &    n.a  &     2.09$\pm$0.26  &     n.a     \\
23 & {\bf NGC 4490} &565&  20501578&  0.430$\pm$0.014 & 0.333$\pm$0.017 &  $<$0.0063 & 1.29$\pm$0.11 & n.a\\
&&&&  0.423$\pm$0.084 &  0.328$\pm$0.065 &  0.011$\pm$0.002 & 1.29$\pm$0.51 & 29.81$\pm$11.33 \\
24 & {\bf NGC 4670} &1609&  58000205&  0.0954$\pm$0.0083 & 0.717$\pm$0.075 & 0.0248$\pm$0.0051 & 0.13$\pm$0.03 & 28.91$\pm$8.97 &\\
&&&&  0.094$\pm$0.008     &  $<$1.13      &     $<$0.17  &     n.a &     n.a   \\
25 & {\bf NGC 4945} &563&  28000446&  3.547$\pm$0.075 & 1.82$\pm$0.061 & 0.38$\pm$ 0.021 & 1.95$\pm$0.11 & 4.79$\pm$0.43 &\\
&&&&  3.52$\pm$0.70 &    1.93$\pm$0.39 &    0.34$\pm$0.07 & 1.82$\pm$0.73 & 5.67$\pm$2.31 \\
26 & {\bf Cen A} &547&  63400464&  2.764$\pm$0.085 & 1.757$\pm$0.068 & 0.102$\pm$0.0062 & 1.57$\pm$0.11 & 17.23$\pm$1.71 &\\
&&&&  2.90$\pm$0.58 &    1.92$\pm$0.39 &    0.10$\pm$0.02 & 1.51$\pm$0.61 & 19.2$\pm$7.74\\
27 & {\bf NW Cen A} &547&  45400151&  2.79$\pm$0.22 & 0.984$\pm$0.070 & 0.0744$\pm$0.0058 & 2.83$\pm$0.43 & 13.22$\pm$1.97 &\\
&&&&  2.43$\pm$0.48 &    0.90$\pm$0.18 &   0.08$\pm$0.02 & 2.7$\pm$1.1 & 11.25$\pm$5.06 \\
28 & {\bf M 51} &600&  35100651&  0.951$\pm$0.048 & 0.623$\pm$0.057 & 0.0268$\pm$0.0038 & 1.53$\pm$0.22 & 23.25$\pm$5.42 &\\
&&&&  1.04$\pm$0.01 &    0.44$\pm$0.09 &      n.a   & 2.36$\pm$0.51 & n.a     \\
29 & {\bf M 83} &513&  64200513&  1.202$\pm$0.051 &  1.37$\pm$0.29 & 0.137$\pm$0.032 & 0.88$\pm$0.22 & 10.0$\pm$4.45 &\\
&&&&  1.76$\pm$0.35 &    1.18$\pm$0.24 &    0.10$\pm$0.02 & 1.49$\pm$0.60 & 11.8$\pm$4.8 \\
30 & {\bf Circinus} &434&  10401133&  2.61$\pm$0.10 & 2.170$\pm$0.084 & 0.1606$\pm$0.023 & 1.20$\pm$0.09 & 3.58$\pm$0.27 &\\
&&&&  2.65$\pm$0.53 &    2.30$\pm$0.46 &    0.18$\pm$0.05 & 1.15$\pm$0.46 & 12.77$\pm$6.10 \\
31 & {\bf Mrk 297} &4739&  62702069&  0.227$\pm$0.027 & 0.291$\pm$0.037 & $<$0.19 & 0.78$\pm$0.19 & n.a\\
&&&&  0.21$\pm$0.008  &        0.225$\pm$0.18     &        $<$0.18  &     0.93$\pm$0.78 &     n.a   \\
32 & {\bf NGC 6240} &7339&  27801108&  0.2447$\pm$0.0082 & 0.651$\pm$0.027 & 0.0349$\pm$0.0029 & 0.38$\pm$0.03 & 18.65$\pm$2.32 &\\
&&&&  0.29$\pm$0.06 &    0.69$\pm$0.14 &  0.031$\pm$0.007 & 0.42$\pm$0.17 & 22.26$\pm$9.54 \\
33 & {\bf NGC 6810} &2031&  84700610& 0.311$\pm$0.014 & 0.22$\pm$0.031 &  $<$0.016 & 1.41$\pm$0.26  & n.a\\
&&&&  0.40$\pm$0.08 &    0.18$\pm$0.04 &     n.a   & 2.22$\pm$0.93 & n.a  \\
34 & {\bf NGC 6946} &48&  45700139&  0.876$\pm$0.028 & 0.611$\pm$0.024 &  $<$0.061 & 1.43$\pm$0.10 & n.a\\
&&&&  1.03$\pm$0.21 &    0.59$\pm$0.12 &    0.05$\pm$0.01 & 1.74$\pm$0.71 & 11.8$\pm$4.8 \\
\end{tabular}
}
\end{minipage}
\end{table*}
\begin{table*}
\begin{minipage}{180mm}
\scriptsize{
\begin{tabular}{llllllllll}
\hline
{\bf Index}& {\bf Source}&{\bf Radial Vel}&{\bf TDT No. of }&{\bf [CII] 158$\mu$m}&{\bf [OI] 63$\mu$m}&{\bf [OI] 145$\mu$m}& {\bf \underline{[CII]$_{158}$}} & {\bf \underline{[OI]$_{63}$}}\\
& & {\bf (km s$^{-1}$)} &{\bf Observation }& & & &{\bf [OI]$_{63}$} &{\bf [OI]$_{145}$}\\
\hline
35 & {\bf NGC 7673} &3408&  76601364&  0.1061$\pm$0.0041 & 0.172$\pm$0.041 &  $<$0.0075 & 0.62$\pm$0.17 & n.a\\
&&&&  n.a  &         n.a     &         n.a  &     n.a &     n.a   \\
36 & {\bf Mrk 331} &5541&  56500637&  0.182$\pm$0.006 & 0.096$\pm$0.008 & 0.0076$\pm$0.0018  & 1.89$\pm$0.22 & 12.63$\pm$4.03\\
&&&&  0.148$\pm$0.008  &         0.11$\pm$0.006     &     $<0.13$  &     1.34$\pm$0.14 &     n.a   \\
37 & {\bf NGC 4151} &995&  35300163&  0.822$\pm$0.006 & 0.445$\pm$0.024 &  $<$0.042 & 1.85$\pm$0.11 & n.a\\
&&&& 0.074$\pm$0.006  &    0.37$\pm$0.029     &         n.a  &     0.20$\pm$0.03 &     n.a   \\
38 & {\bf NGC 6286} &5501&  20700509&  0.187$\pm$0.008 & 0.087$\pm$0.004 &  0.0094$\pm$0.0013 & 2.15$\pm$0.19 & 9.25$\pm$1.70\\
&&&&  0.168$\pm$0.005  &         0.073$\pm$0.009     &       0.005$\pm$0.001  &     2.30$\pm$0.35 &     14.6$\pm$4.7   \\
39 & {\bf NGC 6574} &2282&  70500604&  0.436$\pm$0.021 & 0.223$\pm$0.034 &  0.078$\pm$0.017 & 1.95$\pm$0.39 & 2.86$\pm$1.05\\
&&&&  0.44$\pm$0.020  &         0.23$\pm$0.041     &         n.a  &     1.91$\pm$0.42 &     n.a   \\
40 & {\bf NGC 6822} &-57&  34300915&  0.176$\pm$0.011 & 0.128$\pm$0.005 & 0.008$\pm$0.002 & 1.37$\pm$0.15 & 16.5$\pm$3.9\\
&&&&   0.187$\pm$0.011 &      0.143$\pm$0.006     &    $<$0.05  &     1.30$\pm$0.13 &     n.a   \\
41 & {\bf NGC 7552} &1608&  36903087&  0.583$\pm$0.028 & 0.20$\pm$0.04 & 0.037$\pm$0.007  & 2.92$\pm$0.72 & 5.41$\pm$2.10\\
&&&& 0.64$\pm$0.015  &    0.63$\pm$0.023     &   n.a  &     1.01$\pm$0.06 &     n.a   \\
42 & {\bf NGC 7771} &4277&  56500772&  0.29$\pm$0.01 & 0.141$\pm$0.009 &  0.0187$\pm$0.0026 & 2.06$\pm$0.20 & 7.54$\pm$1.52\\
&&&&  0.298$\pm$0.009  &      0.115$\pm$0.013     &   n.a  &   2.59$\pm$0.37 &     n.a   \\
43 & {\bf NGC 4041} &1234&  22202506&  0.32$\pm$0.01 & 0.38$\pm$0.05 & $<$0.14  & 0.84$\pm$0.13 & n.a\\
&&&&  0.348$\pm$0.005  &        0.197$\pm$0.016     &      n.a  &     1.76$\pm$0.16 &     n.a   \\
44 & {\bf NGC 0278} &627&  59702260&  0.72$\pm$0.002 & $<$0.03 &  0.020$\pm$0.002 & n.a & n.a\\
&&&&  0.728$\pm$0.012 &       0.347$\pm$0.016     &       $<$0.021    &   2.09$\pm$0.13 &     n.a   \\
45 & {\bf NGC 0695} &9735&  63300744&  0.137$\pm$0.007 & 0.19$\pm$0.01 &  0.010$\pm$0.002 & 0.72$\pm$0.07 & 19$\pm$4.8\\
&&&&  0.20$\pm$0.008  &        0.118$\pm$0.009    &         n.a  &     1.69$\pm$0.19 &     n.a   \\
46 & {\bf NGC 0986} &1974&  74300187&  0.278$\pm$0.009 & 0.19$\pm$0.04 &  0.013$\pm$0.001 & 1.46$\pm$0.35 & 14.62$\pm$4.20\\
&&&&  0.304$\pm$0.011  &         0.127$\pm$0.013     &      $<$0.025  &     2.39$\pm$0.33 &     n.a   \\
\end{tabular}
}
\end{minipage}
\end{table*}

Table~\ref{sources} lists the emission line fluxes measured by us from the 
ISO spectra. For 33 of the galaxies, fluxes previously published in the 
literature were available and these are also listed in 
Table~\ref{sources}. When uncertainties were not listed for these 
literature fluxes, we have estimated them as 20$\%$ for fluxes $\leq$ 
10$^{-14}$ W m$^{-2}$ and 10$\%$ for fluxes $\geq$ 10$^{-14}$ W m$^{-2}$. 
Although the choice of how to estimate the associated errors, when it 
is not provided, is arbitrary, we used typical 
uncertainties resulting from calibration and pointing errors in the 
submillimeter and IR 
domain \citep{Martin06,Bayet04,Israel95}. The percentage uncertainty
estimated in this way appears, on average, to be similar to the percentage 
line flux uncertainties measured by us using the interactive package 
SMART.

\begin{figure}
\psfig{file=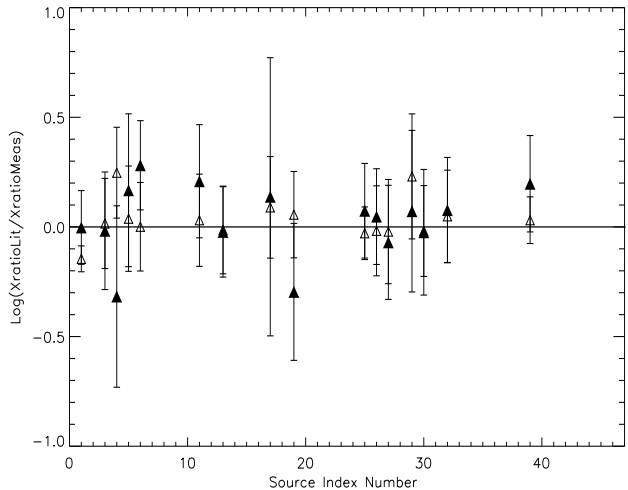,height=7cm,width=8.81cm}
\caption{[C~{\sc ii}]$_{158}$ /[O~{\sc i}]$_{63}$ and [O~{\sc 
i}]$_{63}$/[O~{\sc i}]$_{145}$ ratios. Filled triangles represent the 
ratio of the literature value of [O~{\sc i}]$_{63}$/[O~{\sc i}]$_{145}$ 
over our own measured [O~{\sc i}]$_{63}$/[O~{\sc i}]$_{145}$ ratio, with 
respective error bars. Open triangles represent the ratio of the 
literature value of [C~{\sc ii}]$_{158}$ /[O~{\sc i}]$_{63}$ over our own 
measured [C~{\sc ii}]$_{158}$ /[O~{\sc i}]$_{63}$ ratio with respective 
error bars. The horizontal line represents the case in which literature 
ratios and our own measured ratios are perfectly matched. The source 
index numbers and galaxy identifications are listed in Table~\ref{sources}.}
\label{one}
\end{figure}

In order to show that our measurements are in good agreement with the 
fluxes collected from the literature we have plotted our measurements 
versus literature values in Fig~\ref{one}. We have plotted the ratio of 
the literature value of [C~{\sc ii}]$_{158}$ /[O~{\sc i}]$_{63}$ over our 
own measured [C~{\sc ii}]$_{158}$ /[O~{\sc i}]$_{63}$ value, with 
respective error bars (open triangles), as well as the ratio of the 
literature value of [O~{\sc i}]$_{63}$/[O~{\sc i}]$_{145}$ over our own 
value of [O~{\sc i}]$_{63}$/[O~{\sc i}]$_{145}$, with respective error 
bars (filled triangles). The horizontal line in Fig~\ref{one} represents 
the case in which the literature ratios and our own measured ratios are in 
perfect agreement.

Flux ratios of two lines of the same species, such as [O~{\sc 
i}]$_{145}$/[O~{\sc i}]$_{63}$ can provide information about nebular 
conditions such as temperature (or radiation field strength).
In addition, a 
ratio such as [O~{\sc iii}]$_{88}$/[O~{\sc iii}]$_{52}$ can provide 
information on the electron density. In addition by using ratios, rather 
than fitting the line intensities directly, the beam filling factors of 
the two emission lines cancel out, assuming that they come from the same 
regions and are the same size.

\begin{figure}
\psfig{file=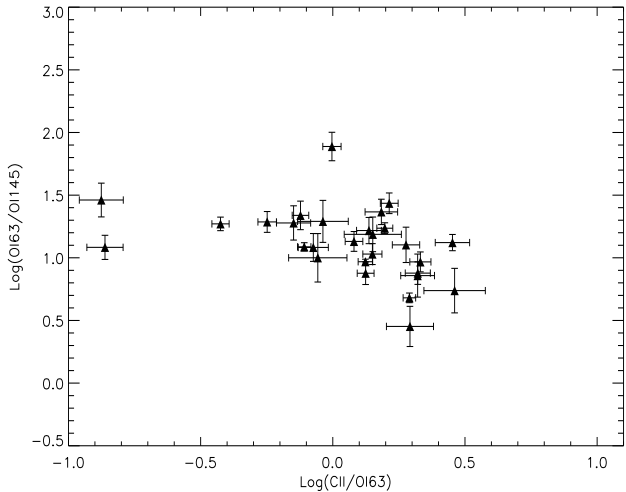,height=7cm,width=8.81cm}
\caption{The observed [C~{\sc ii}]$_{158}$ /[O~{\sc i}]$_{63}$ ratios versus the [O~{\sc i}]$_{63}$/[O~{\sc i}]$_{145}$ ratios for 
detections $\geq$ 4$\sigma$, with error bars.}
\label{due}
\end{figure}

Fig~\ref{due} shows the [C~{\sc ii}]$_{158}$ /[O~{\sc i}]$_{63}$ ratio 
versus the [O~{\sc i}]$_{63}$/[O~{\sc i}]$_{145}$ ratio for the 28 
sources, out of the 46 listed in Table~\ref{sources}, that showed 
$\geq$ 4$\sigma$ detections in all three lines.

\section{The UCL\_PDR models}
\label{sec:Model}
We used the {\bfseries UCL\_PDR} time and depth-dependent PDR code 
\citep{Bell:06b} which solves the chemistry, thermal balance and radiative 
transfer self-consistently within a cloud. The range of parameter values 
input to the code are listed in Table~\ref{params}, including the mean
grain radius and grain number density adopted for grain surface reactions.
The FUV grain opacity properties adopted by UCL\_PDR are listed by \citet{Bell:06}. The outputs of the 
code are the abundances of each species included in the chemical network, 
their column densities, the emissivities and integrated intensities of the 
emission lines involved in the cooling, and the gas and dust temperatures.
All of these quantities are functions of depth and time. At each 
depth-step, the model calculates the attenuation of the FUV field before 
beginning an iterative cycle to determine the gas temperature at which the 
total heating and cooling rates are equal, that is, when the condition of 
thermal balance is satisfied. For each iteration, the chemistry is first 
calculated, based on the gas temperature and attenuated FUV flux, after 
which the heating and cooling rates are computed, together with the 
radiative transfer in the emission line, using the revised chemical 
abundances. The difference between the total heating and cooling rates is 
then used to derive a new gas temperature. We use a grid of 1702 PDR 
models, already computed and partially used in previous works 
(Bell at al. 2006, Bell $\&$ Hartquist et al. 2006, Bell et al. 2007) spanning a large range of densities, 
metallicities and cosmic ray ionization rates. Amongst the range of 
parameter values covered by the grid of models (see Table~\ref{params}), 
we do not consider variations of the turbulent velocity parameter in our 
attempts to fit the observed emission from each galaxy, since the line 
ratios are believed to be fairly insensitive to small changes in the 
turbulent velocity \citep{Wolf89}.

\begin{table}
\scriptsize{
\caption{Physical and chemical parameters for the UCL\_PDR grid. In the grid of models the 
density and the radiation field parameters are incremented by 2 dex
and 1 dex, respectively.} \label{params}
\begin{tabular}{lll}
\hline
\bfseries Parameter &\bfseries Range of values & \bfseries Reference\\
& &\bfseries  parameter\\
& &\bfseries  values\\
\hline
Cloud density (cm $^{-3}$)& 10$^{2}$$\leq$ n$_H$ $\leq$10$^{5}$ & n$_H$= 10$^{3}$\\
\hline
Incident FUV flux (Habing)& 10$\leq$ G$_{0}$$\leq$10$^{5}$  & G$_{o}$=10\\
\hline
Age of the cloud (yr)& 10$^{4}$$\leq$ t $\leq$10$^{8}$ & t=10$^{7}$\\
 \hline
Cloud size (mag)& 0$\leq$ A$_{v}$$\leq$10 & A$_{v}$=10\\
\hline
Metallicity& Z/Z$_{\odot}$=5,4,3,2,1,0.5& Z=Z$_{\odot}$ \\
& 0.1,0.25,0.01&\\
\hline
C.R. ionization rate (s$^{-1}$)& $\zeta$ = 5x,50x,500x10$^{-15}$ & $\zeta$= 5x10$^{-17}$\\
&5x,50x,500x10$^{-16}$&\\
&5x,50x,500x10$^{-17}$&\\
\hline
Turbulent velocity (km s$^{-1}$)& v$_{turb}$=1.5 & v$_{turb}$=1.5\\ 
\hline
Carbon elemental abundance & 1.4x10$^{-4}$n$_{H}$ & Sofia  \& Myer \\
&& (2001) \\
\hline
Oxygen elemental abundance & 3.2x10$^{-4}$n$_{H}$& Sofia  \& Myer \\
&& (2001) \\
\hline
Grain radius & 0.1~$\mu$m & 0.1~$\mu$m \\ 
\hline
Grain number density n$_{g}$ & 2x10$^{-12}$Z $n$ cm$^{-3}$ &2x10$^{-9}$ cm$^{-3}$\\

\end{tabular}
}
\end{table}

\section{Sensitivity of model results to parameter variations}
\label{sec:Trends}

We attempted to find the range of physical parameters which best 
reproduced the observations, such as visual extinction, metallicity, 
cosmic ray ionization rate and UV radiation field strength. The influence 
of changing the number density of hydrogen nuclei (n$_{H}$), the visual 
extinction (A$_{v}$), the incident radiation field strength (G$_{o}$), the 
cosmic ray ionization rate $\zeta$ and the metallicity (Z) is considered 
individually by varying only one of these parameters at a time. The age of 
the modeled regions was set to a value of 10$^7$ yrs, owing to the fact 
that clouds with ages $\geq$10$^7$ yrs do not undergo significant changes 
in their predicted chemical profiles, even though, in some cases, the 
chemistry only reaches its final steady state somewhat later 
\citep{Bell:06}.

When one parameter was varied the remainder were held constant at the 
reference parameter values listed in the final column of 
Table~\ref{params}.

In Figs 3 through 8 we overplot our model predictions against flux ratios 
from the [C~{\sc ii}] 158{$\mu$}m and [O~{\sc i}] 63{$\mu$}m and 
145{$\mu$}m 
ISO LWS fluxes. Table~\ref{ncr} lists these 3 important PDR cooling 
transitions, their wavelengths, upper energy levels $E_{upper}$, and 
critical densities $n_{cr}$. In this section we will only discuss the 
sensitivity of the ratios to changes in the input parameters. The 
comparison with observations will be made in Section 6.

\begin{table}
\scriptsize{
\caption{PDR diagnostic transitions. The critical densities for 
[C~{\sc ii}] and [O~{\sc i}] are for collisions with H \citep{TH85}.}
\label{ncr}
\begin{tabular}{lllll}
\hline
& &\bfseries Wavelength& \bfseries E$_{upper}$/k&\bfseries n$_{cr}$\\
\bfseries Species& \bfseries Transitions&  \bfseries ($\mu$m)& \bfseries (K)&\bfseries (cm$^{-3}$)\\
\hline
[C {\sc ii}]& $^2P_{3/2}\-- ^2P_{1/2}$&157.74&92&3x10$^{3}$\\
\hline
[O {\sc i}]& $^3P_{1}\-- ^3P_{2}$&63.18&228&4.7x10$^{5}$\\
 \hline
[O {\sc i}]& $^3P_{0}\-- ^3P_{1}$&145.53&326&1x10$^{5}$\\
\end{tabular}
}
\end{table}

\subsubsection {Visual Extinction}
PDRs derive their properties primarily from the penetration of FUV 
photons 
into their interiors. Dust grains provide the major source of continuum opacity and attenuate the incident radiation field by selectively absorbing and scattering light at visible and ultraviolet wavelengths.
This means that there is a strong correlation between visual extinction and the chemistry of PDRs.
We examined models for different visual extinction values in the range 1$\leq$A$_{v}$$\leq$10.
\begin{figure*}
\psfig{file=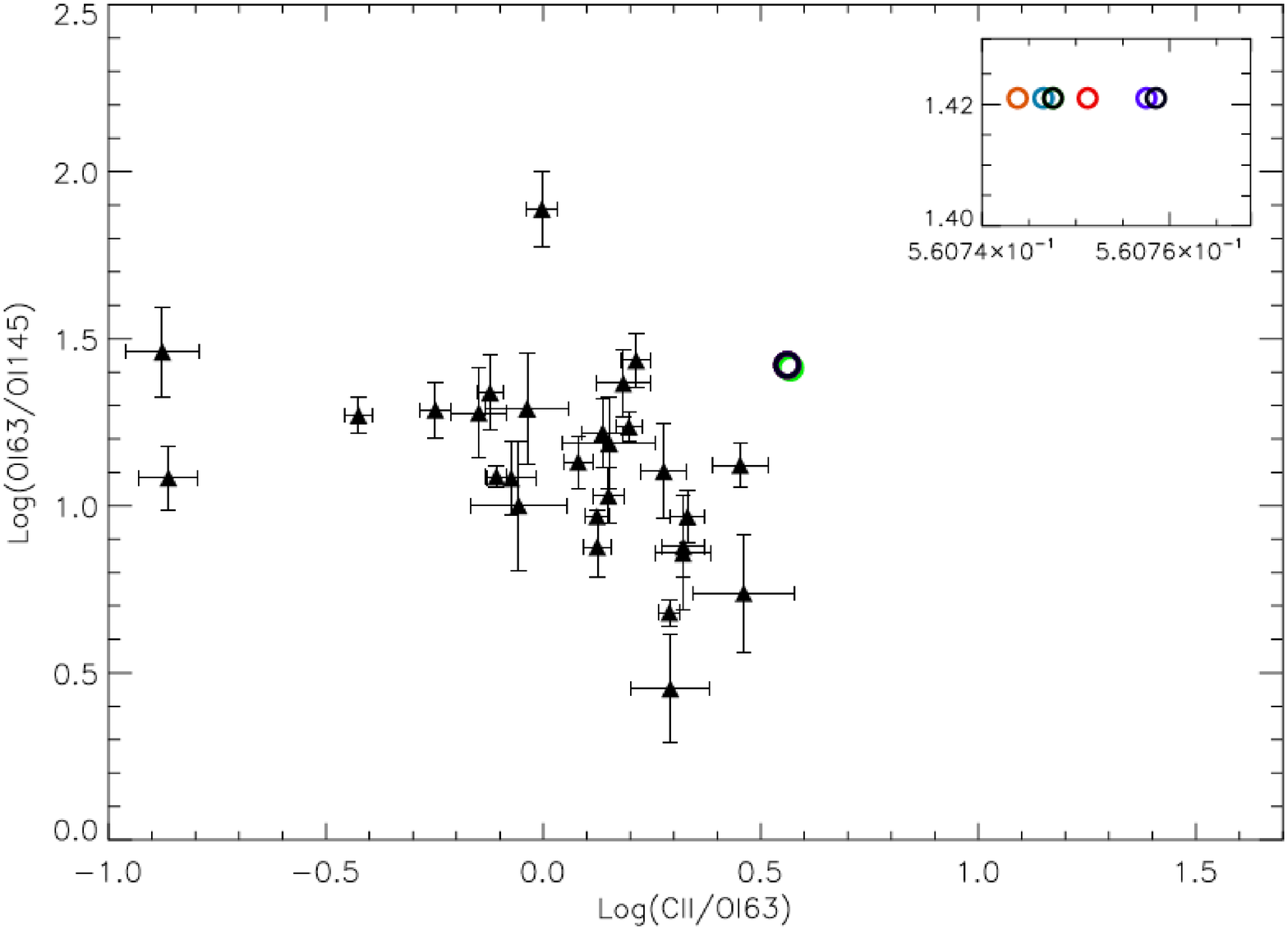,height=7cm,width=8.81cm}
\psfig{file=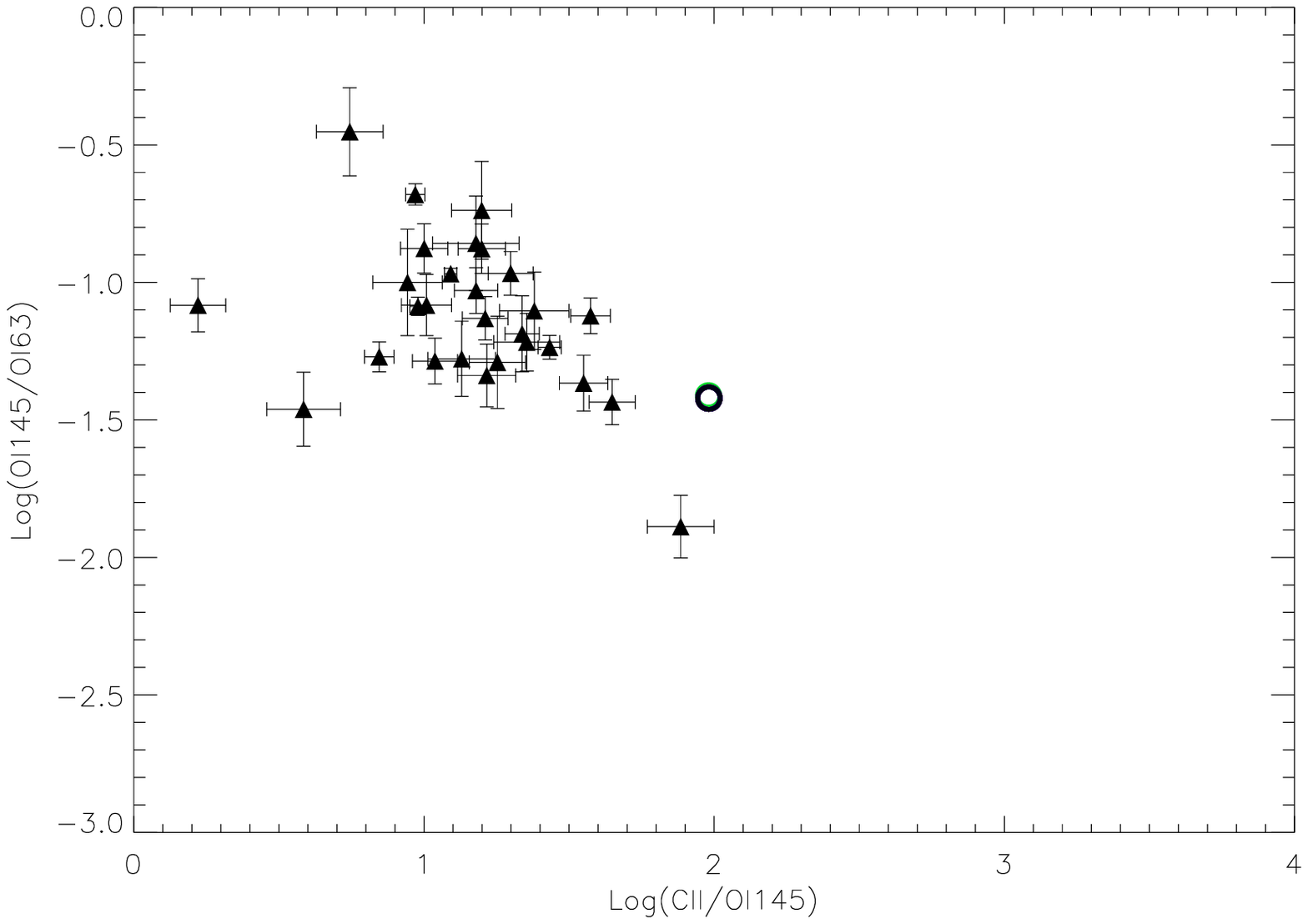,height=7cm,width=8.81cm}
\caption{[C~{\sc ii}]$_{158}$ /[O~{\sc i}]$_{63}$ versus [O~{\sc 
i}]$_{63}$/[O~{\sc i}]$_{145}$ ratios (left) and 
[C~{\sc ii}]$_{158}$ /[O~{\sc i}]$_{145}$ versus 
[O {\sc i}]$_{145}$/[O {\sc i}]$_{63}$ ratios 
(right), with error bars, compared with models of varying visual 
extinctions. The insert, in the left panel, shows a zoomed-in region of 
the overlapping coloured circles corresponding to differing 
visual extinction values.}
\label{tre}
\end{figure*}
From Fig~\ref{tre} one can see that there is no significant change in the 
emission line ratios from A$_{v}$=1 (black circle) to A$_{v}$=10 (green 
circle). This trend implies that the majority of the line fluxes arise 
from 1$\leq$A$_{v}$$\leq$2 and is consistent with the C$^{+}$ abundance 
because, for values of A$_{v}$ larger than 2, this ion recombines to form 
C$^{0}$ and subsequently forms CO. Indeed C$^{+}$ is a tracer of the edge 
of PDRs. This implies that the C$^{+}$ fine structure emission lines that 
are observed are coming predominantly from regions with low visual 
extinction. Although neutral oxygen is still present at higher visual 
extinctions, it is somewhat reduced beyond A$_{v}$=1 because a fraction of 
oxygen combines with C to form CO \citep[see][]{Roellig07}. We fixed 
A$_{v}$=10 as the most appropriate depth range to fit our sample of 
sources, since Fig~\ref{tre} shows that we cannot distinguish between 
models with optical depths in the range of 1$\leq$A$_{v}$$\leq$10.

\subsubsection{Radiation field}

\begin{figure*}
\psfig{file=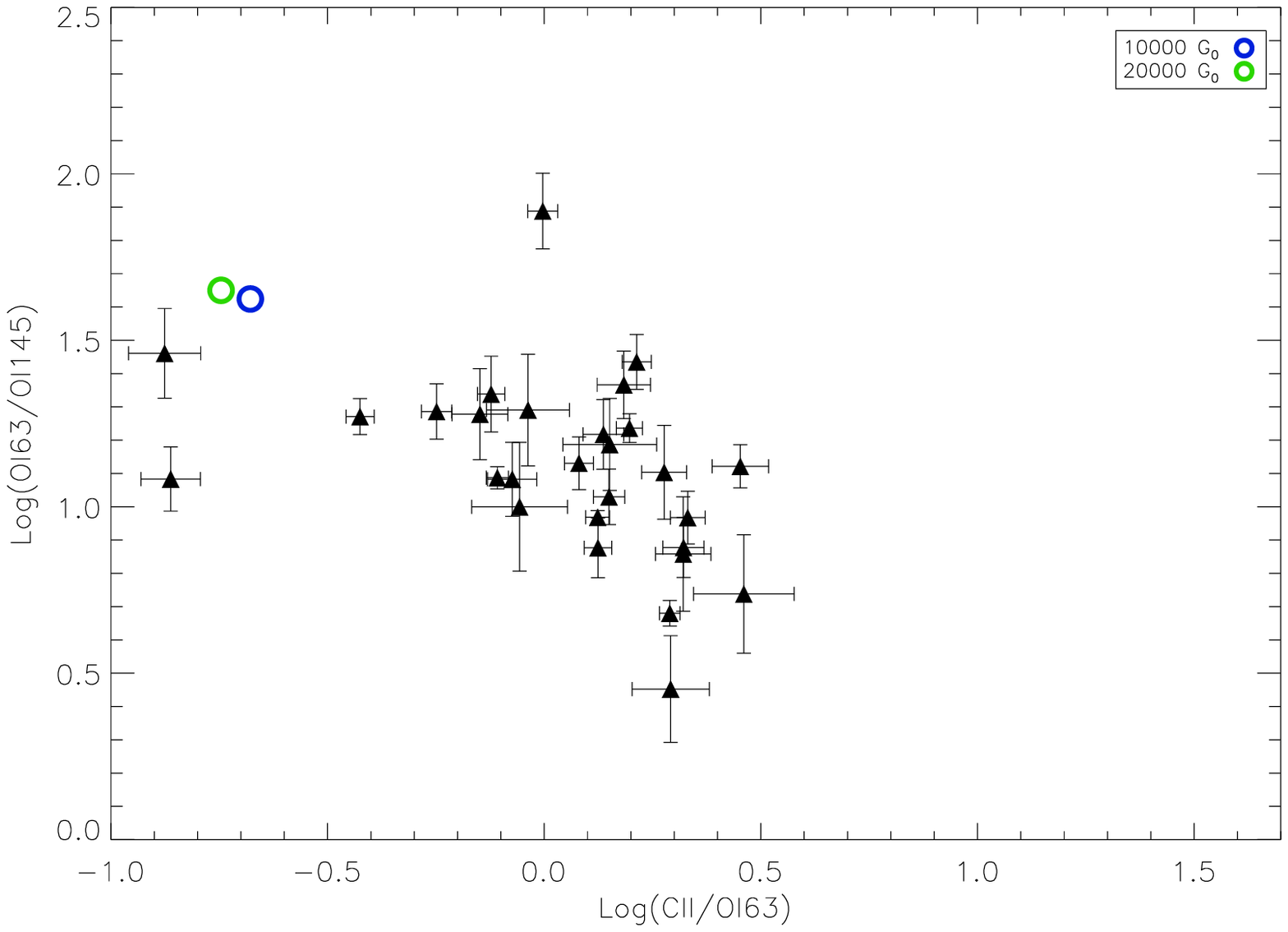,height=7cm,width=8.81cm}
\psfig{file=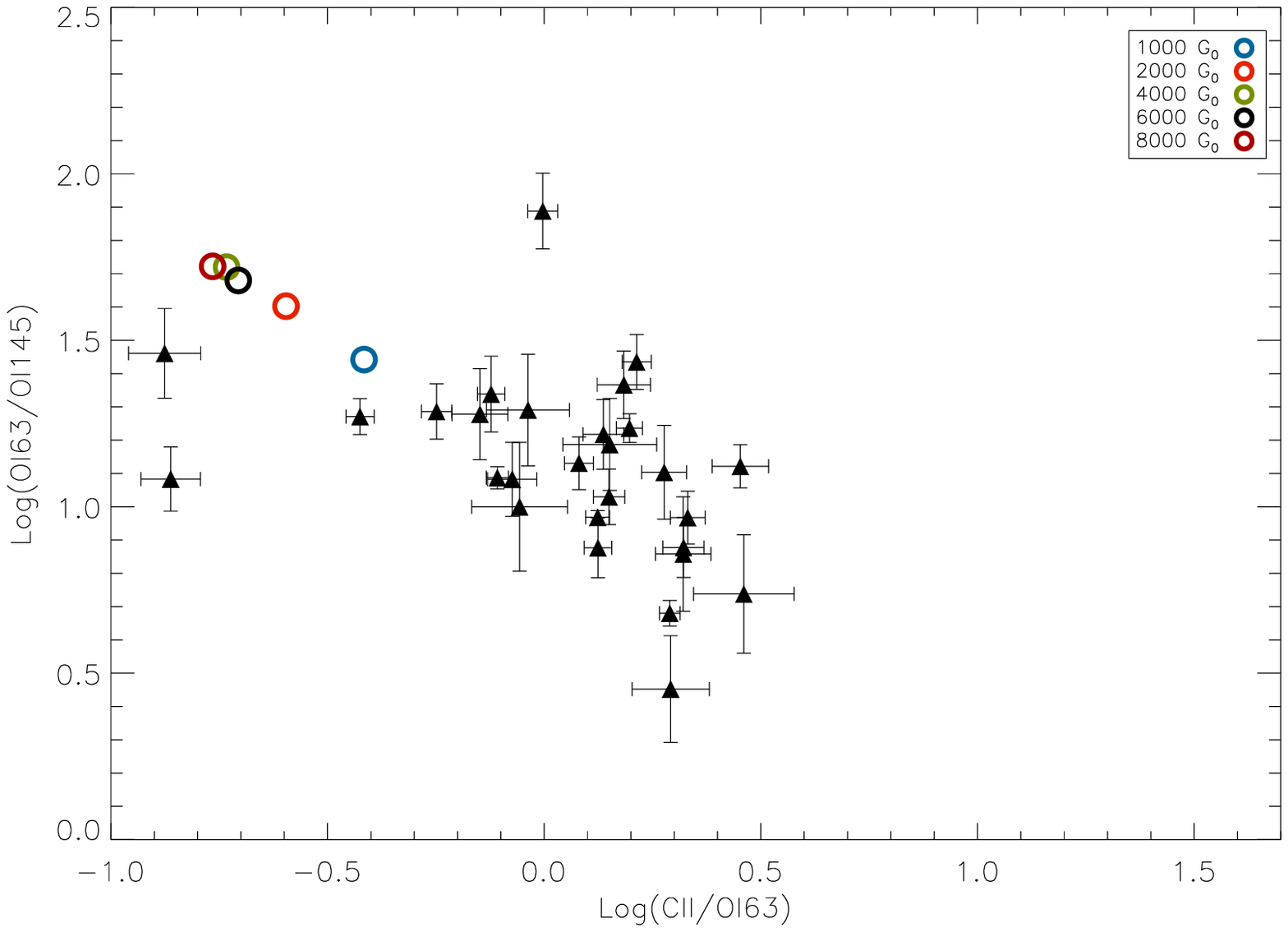,height=7cm,width=8.81cm}
\psfig{file=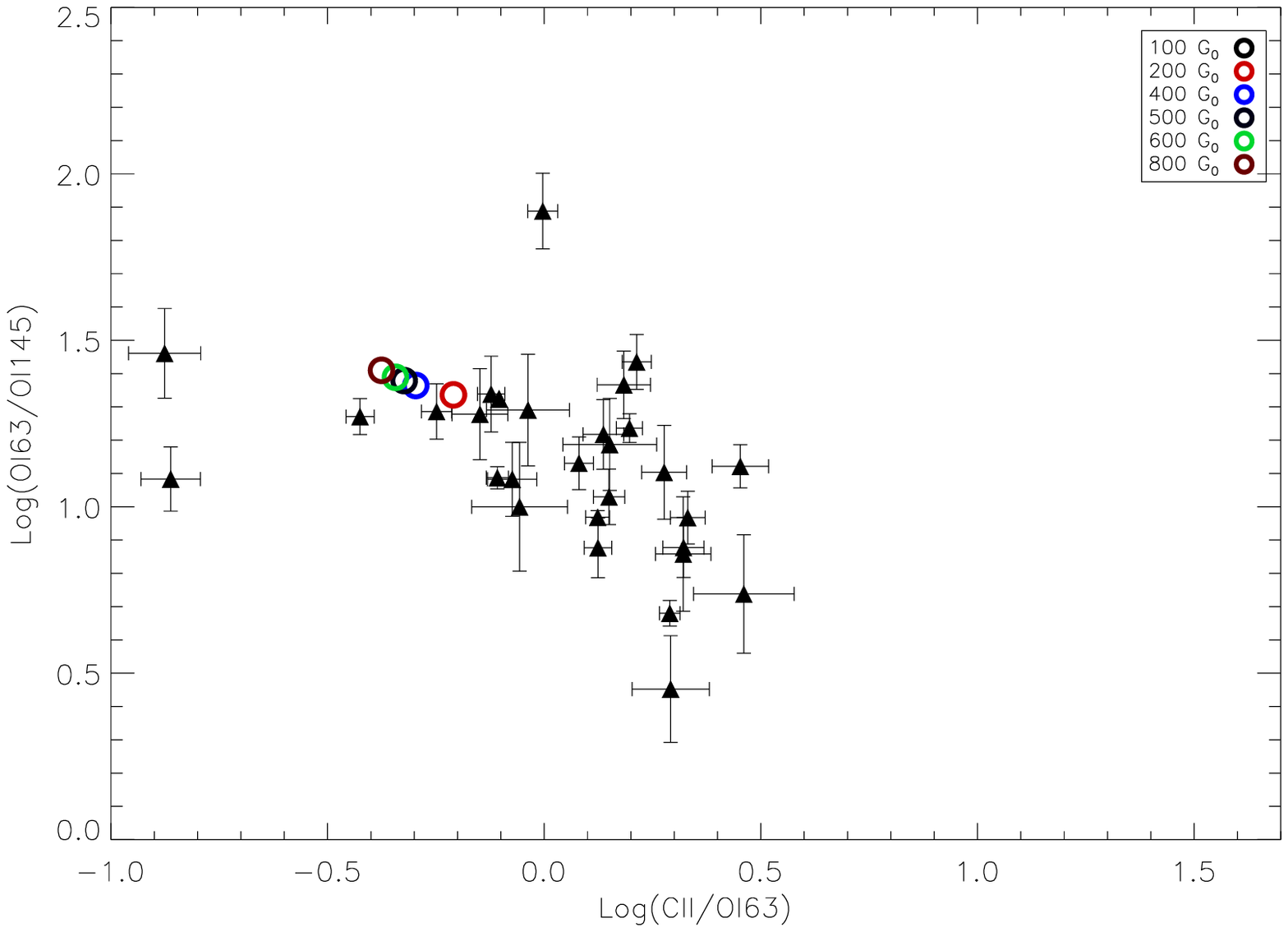,height=7cm,width=8.81cm}
\psfig{file=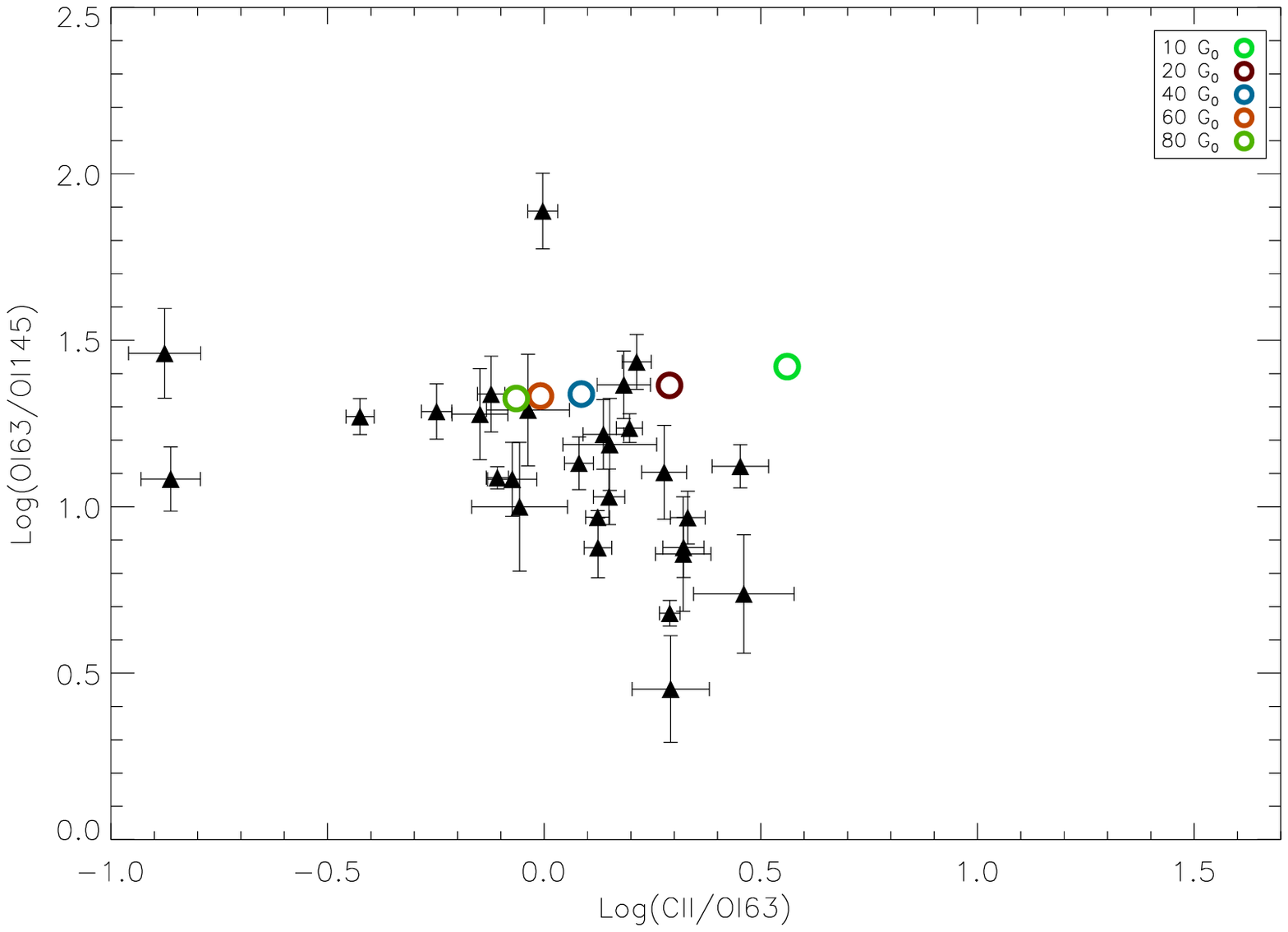,height=7cm,width=8.81cm}
\caption{The four panels show observed [C {\sc ii}]$_{158}$ /[O {\sc i}]$_{63}$ versus [O {\sc i}]$_{63}$/[O {\sc i}]$_{145}$ ratios, with uncertainties. In the top left panel the coloured circles represent models with radiation field strengths in the range 10$^{4}$$\leq$G$_{0}$$\leq$2x10$^{4}$. In the top right panel the coloured circles represent models with radiation field strengths in the range 10$^{3}$$\leq$G$_{0}$$\leq$8x10$^{3}$. In the bottom left panel the coloured circles represent models with radiation field strengths in the range 10$^{2}$$\leq$G$_{0}$$\leq$8x10$^{2}$. In the bottom right panel the coloured circles represent models with radiation field strengths in the range 10$\leq$G$_{0}$$\leq$80.}
\label{quattro}
\end{figure*}

Radiation field strengths incident upon a gas cloud can vary from the 
standard interstellar field ( $\chi$= Draine) up to $\chi$$\sim$10$^{7}$ 
Draines \citep{Draine78} in regions of intense star formation 
\citep[see][]{Hollenbach97}. The FUV field may be also expressed in terms 
of the Habing parameter \citep{Habing68}. The FUV flux expressed in this 
way is determined by the parameter G$_{0}$=$\chi$/1.7. We adopt a range of 
10$\leq$G$_{0}$$\leq$500000 Habings for consideration in this study, but 
we only plot the significant results. From Fig~\ref{quattro}, [C {\sc 
ii}]$_{158}$ /[O {\sc i}]$_{63}$ versus [O {\sc i}]$_{63}$/[O {\sc 
i}]$_{145}$ appears to be sensitive to variations in the radiation field 
strength. As the radiation field strength increases so does the [O {\sc 
i}]$_{63}$/[O {\sc i}]$_{145}$ ratio, while the [C {\sc ii}]$_{158}$ /[O 
{\sc i}]$_{63}$ ratio decreases. The energy required to excite the [O 
{\sc i}]$_{63}$ transition is somewhat higher than that for [C {\sc 
ii}]$_{158}$, hence the line intensity ratio [O {\sc i}]$_{63}$/[C {\sc 
ii}]$_{158}$ is expected to increase with gas temperature and hence higher 
FUV flux. PDR models predict that for n$_H$$>$10$^3$ cm$^{-3}$ the line ratio 
will increase with both G$_{0}$ and n$_H$, due to the different critical 
densities of the two transitions \citep{Wolf90}.

\subsubsection {Metallicity}

Metallicities significantly below solar are observed in Local Group 
galaxies, including the Small Magellanic Cloud, as well as in more distant 
dwarf galaxies, with I Zw 18's metallicity of 1/40 being amongst the 
lowest known \citep{Izotov99}. The metallicity-dependence appears in 
several key processes in the UCL\_PDR code. Elemental abundances of all 
metals are assumed to scale linearly with metallicity 
(Z/Z$_{\odot}$); the dust-to-gas mass ratio is also assumed to 
scale linearly with metallicity and takes a standard a value of 10$^{-2}$ 
at Solar metallicity. The formation rate of H$_{2}$ on grain surfaces and 
the grain photoelectric heating rate are assumed to scale linearly with 
metallicity. 
 
 We investigated a metallicity range of 0.01$\leq$ Z/Z$_{\odot}$$\leq$5 for this study, with the adopted solar neighbourhood carbon and oxygen elemental abundances listed in Table~\ref{params}.
Metallicity affects the total abundances of carbon and oxygen bearing species, and hence can influence the chemical and thermal structure of PDRs in galaxies.

\begin{figure*}
\psfig{file=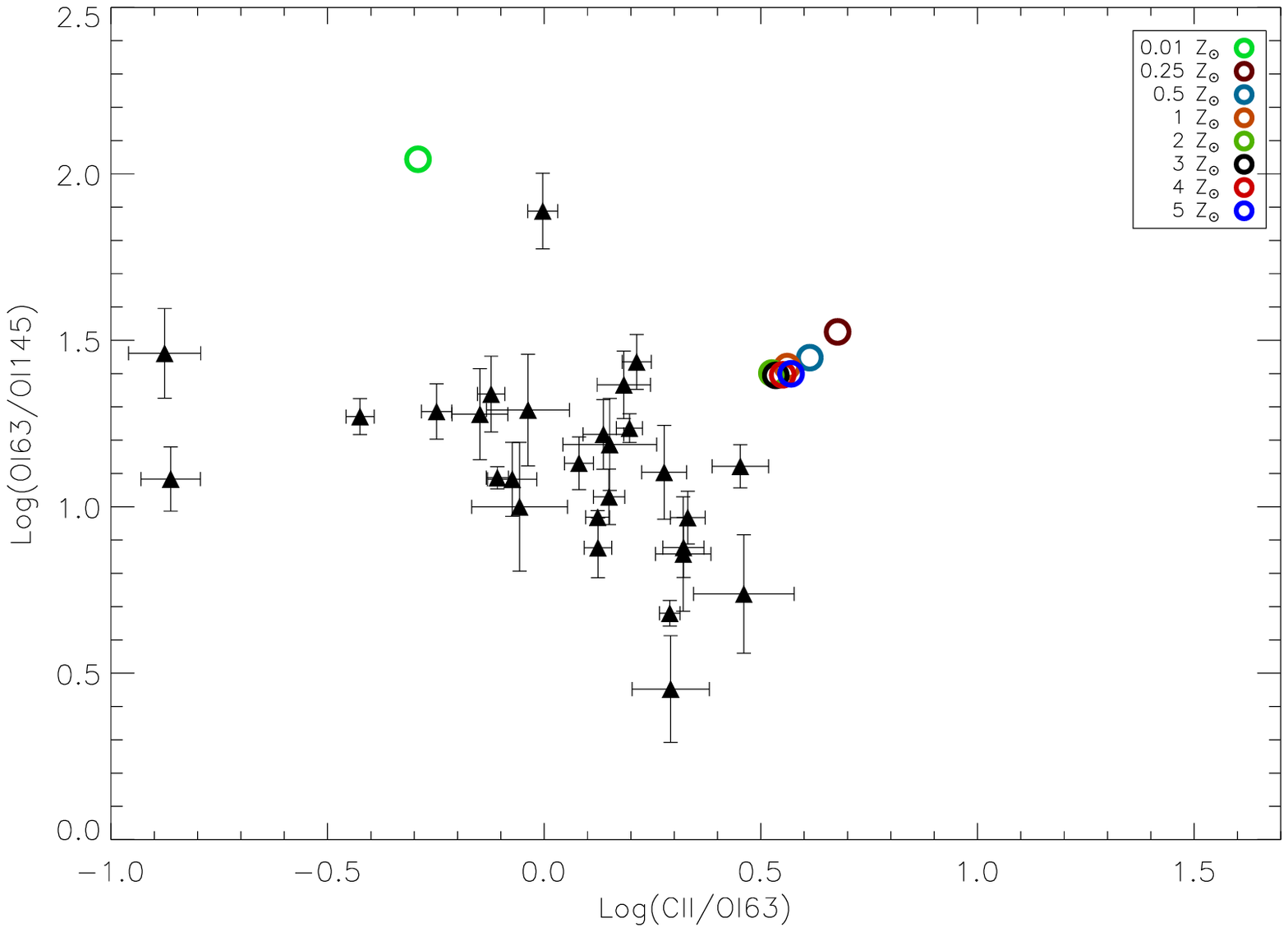,height=7cm,width=8.81cm}
\caption{Observed [C {\sc ii}]$_{158}$ /[O {\sc i}]$_{63}$ versus [O {\sc 
i}]$_{63}$/[O {\sc i}]$_{145}$ ratios, with uncertainties, compared
to models of differing metallicity. The coloured 
circles represent models with different values of metallicity, ranging 
from 0.01 (light green circle), 0.25 (claret circle) and 0.5 (turquoise 
circle) up to 5 times solar metallicity.}

\label{cinque}
\end{figure*}
In Fig~\ref{cinque} we show our model results for metallicities of 0.01 (light green circle), 0.25 (claret circle) and 0.5 (turquoise circle), with remaining symbols representing values of 1, 2, 3, 4 and 5 times solar metallicity respectively.

\begin{figure*}
\psfig{file=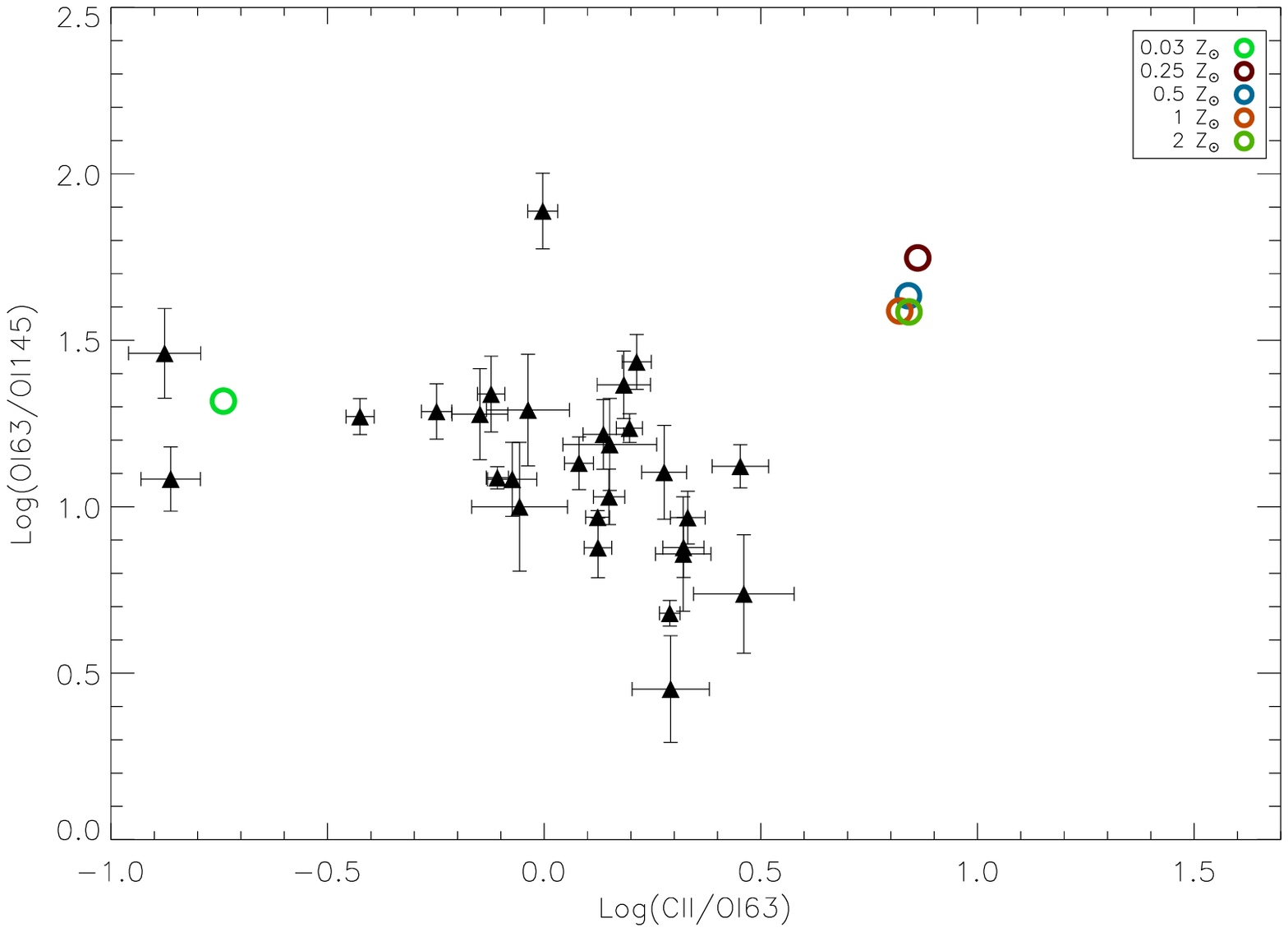,height=7cm,width=8.81cm}
\caption{Observed [C~{\sc ii}]$_{158}$ /[O~{\sc i}]$_{63}$ versus [O~{\sc 
i}]$_{63}$/[O~{\sc i}]$_{145}$ ratios, with uncertainties, compared to
models of differing metallicity, with the C/O ratio adopted to be 
0.53 for metallicities of 0.25 and above, but dropping to 0.16 as
metallicities decrease to 0.03. The coloured 
circles represent models with different values of metallicity, ranging 
from 0.03 (light green circle), 0.25 (claret circle) and 0.5 (turquoise 
circle) up to twice solar metallicity.}

\label{sei}
\end{figure*}

 The C/O ratios for the Sun, the Large Magellanic Cloud (LMC) and the Small Magellanic Cloud (SMC) are all $\sim$0.5 within the uncertainties, but for lower metallicities than the SMC C/H decreases faster than O/H. We tried to reproduce this trend by changing carbon and oxygen elemental abundances with metallicity.
In Fig~\ref{sei} we show our model results for a metallicity of Z/Z$_{\odot}$=1, with C/O=0.53  \citep{Asplund05}, then we scaled all the elemental abundances with Z/Z$_{\odot}$ down to Z/Z$_{\odot}$=0.25. Below Z/Z$_{\odot}$=0.25 the C/O ratio was scaled from 0.53 to 0.16 as Z/Z$_{\odot}$ was reduced from 0.25 to 0.03 \citep{Izotov99}.
As expected, the [O {\sc i}]$_{63}$/[O {\sc i}]$_{145}$ ratio is not sensitive to variations of metallicity whereas the [C {\sc ii}]$_{158}$ /[O {\sc i}]$_{63}$ ratio would appear at first sight to be a good tracer of metallicity.

\subsubsection{Density}

\begin{figure*}
\psfig{file=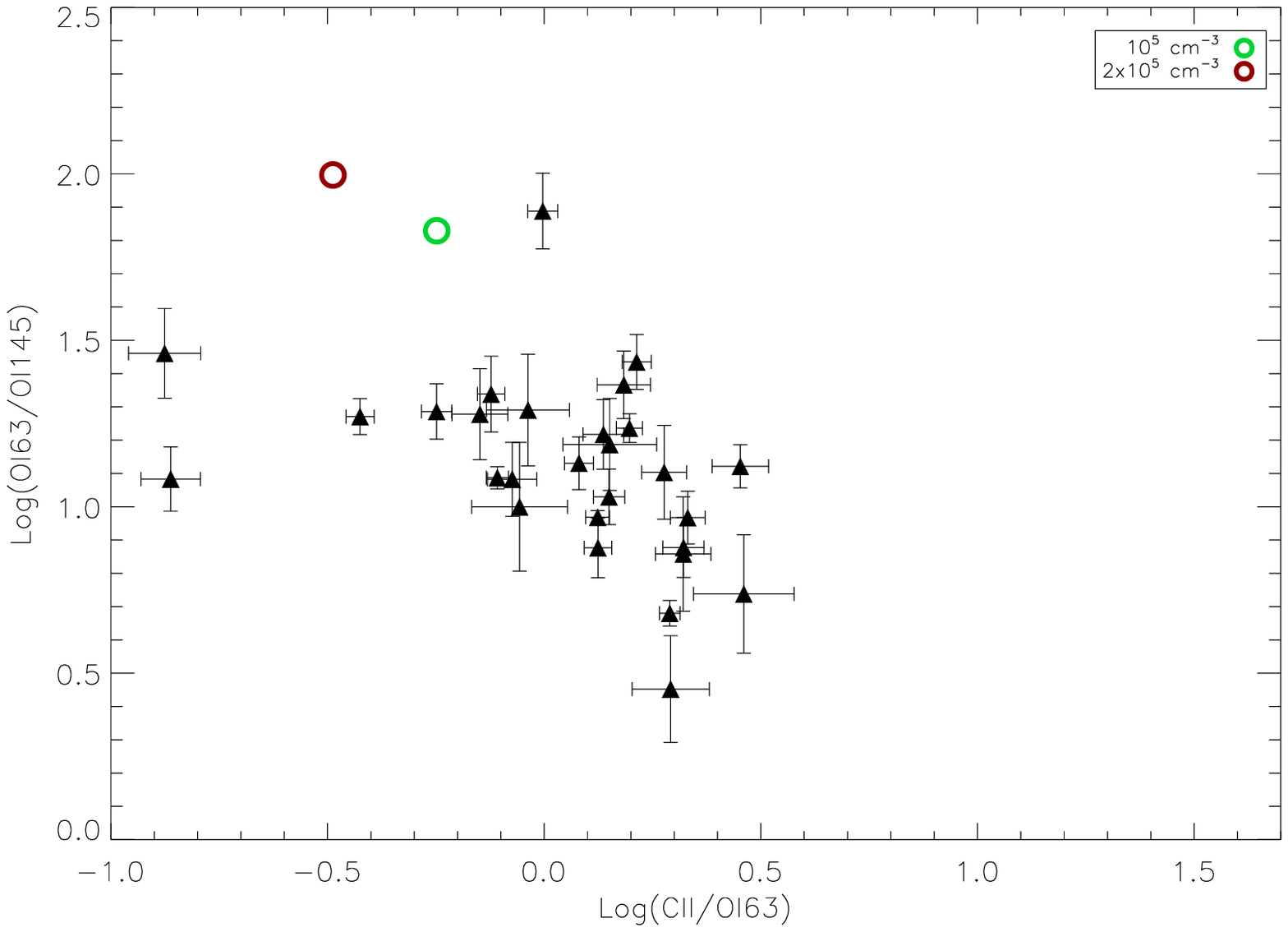,height=7cm,width=8.81cm}
\psfig{file=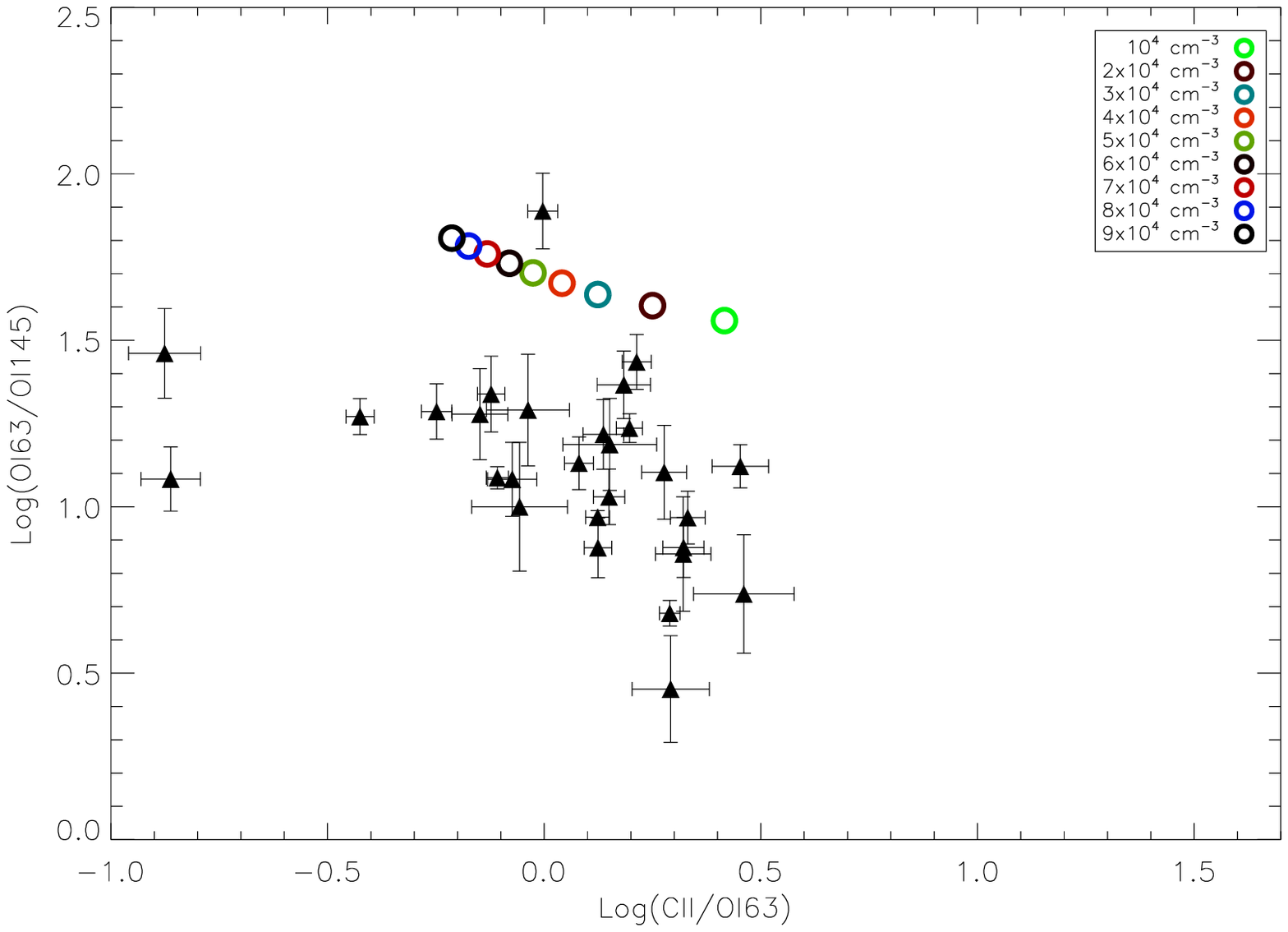,height=7cm,width=8.81cm}
\psfig{file=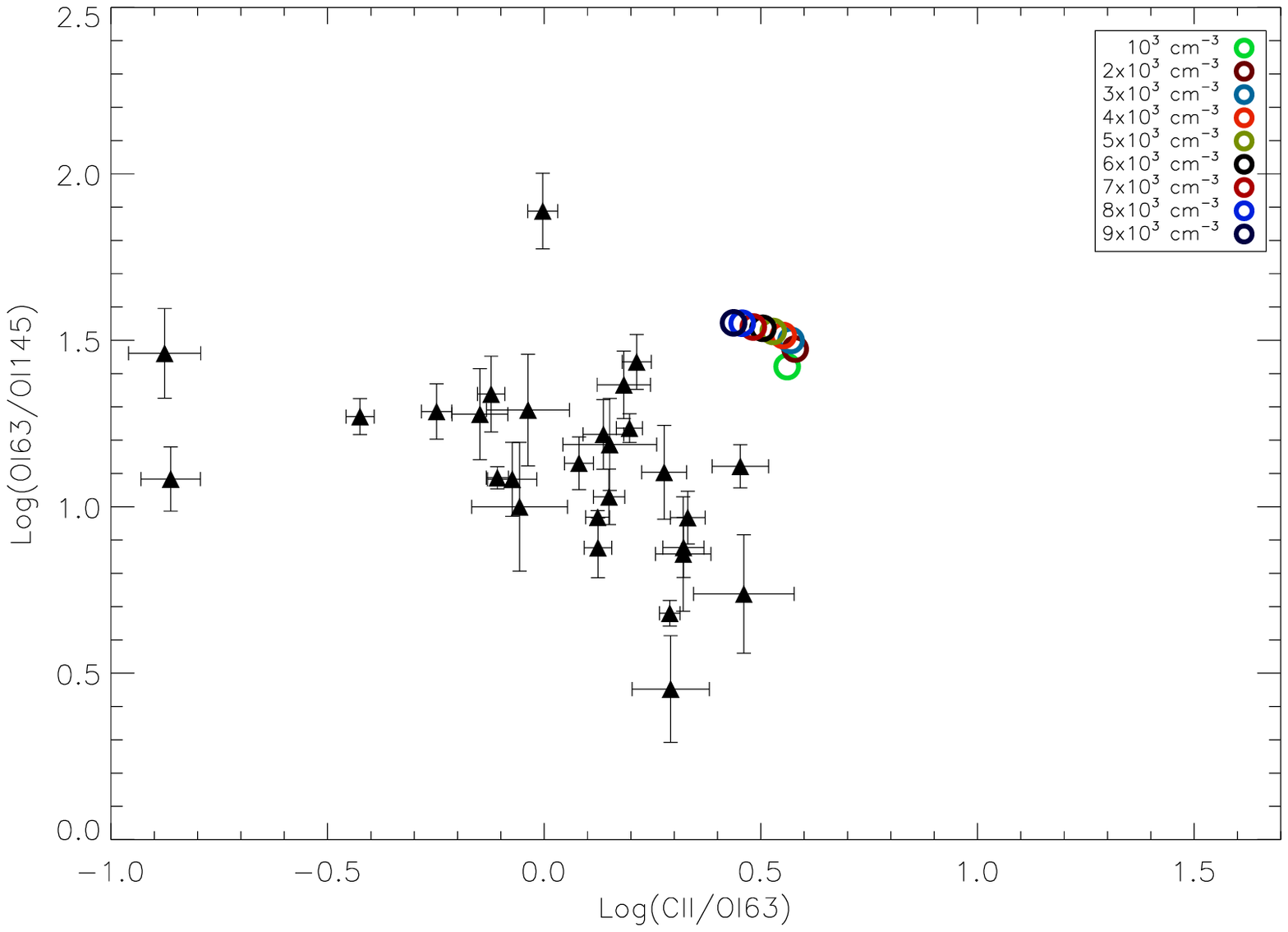,height=7cm,width=8.81cm}
\psfig{file=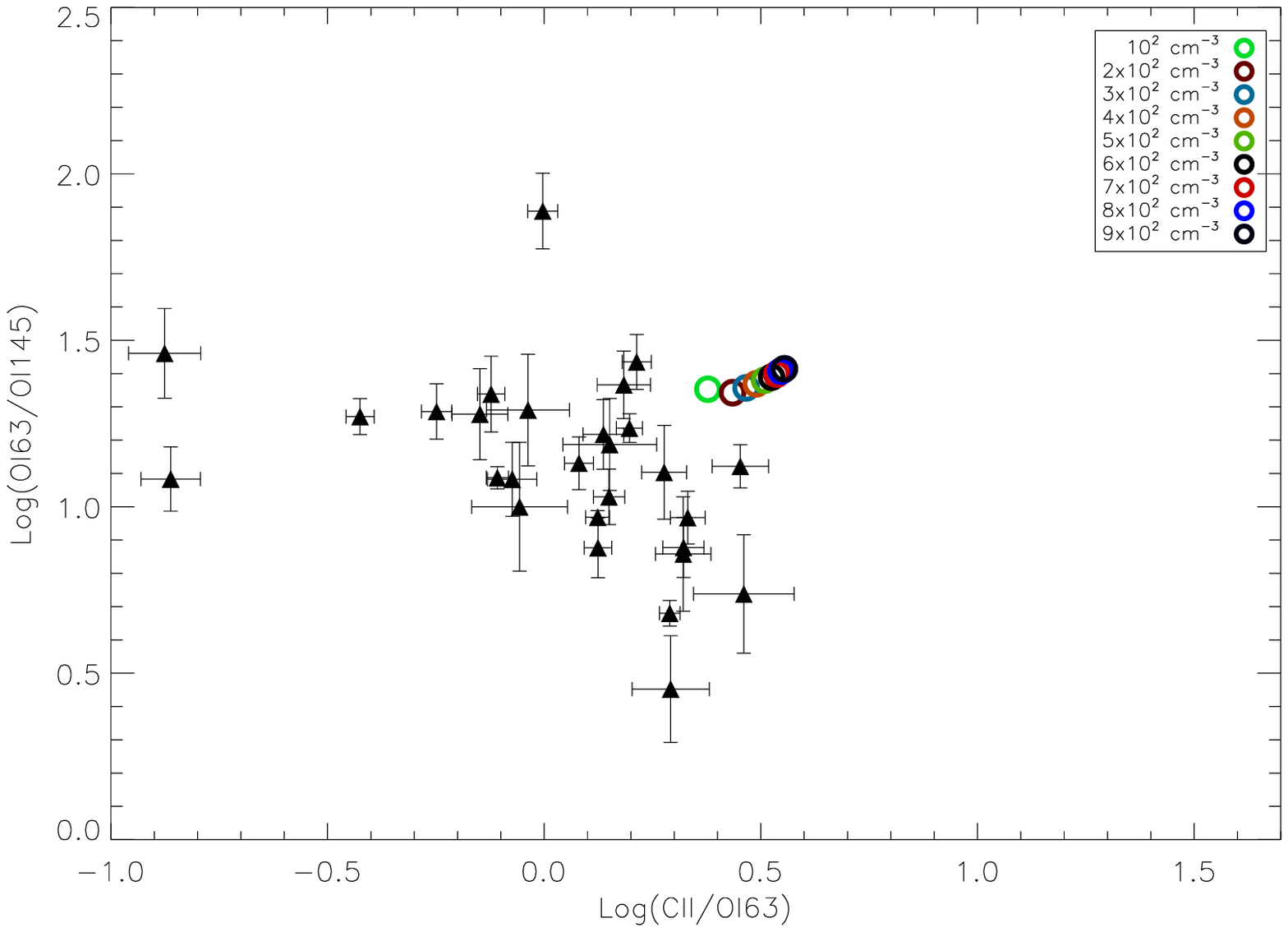,height=7cm,width=8.81cm}
\caption{Observed [C {\sc ii}]$_{158}$ /[O {\sc i}]$_{63}$ versus [O {\sc 
i}]$_{63}$/[O {\sc i}]$_{145}$ ratios, with uncertainties,
compared to models of differing densities. In the left top 
panel the colored points represent models with densities in the range 
10$^{5}$$\leq$ n$_{H}$$\leq$2x10$^{5}$ cm$^{-3}$ decreasing from left to 
right. In the right top panel the colored points represent models with 
densities values in the range 10$^{4}$$\leq$ n$_{H}$$\leq$9x10$^{4}$ 
cm$^{-3}$, decreasing from left to right.In the left bottom panel the 
colored points represent models with densities in the range 10$^{3}$$\leq$ 
n$_{H}$$\leq$9x10$^{3}$ cm$^{-3}$ decreasing from left to right. In the 
right bottom panel the colored points represent models with densities 
values in the range 10$^{2}$$\leq$ n$_{H}$$\leq$9x10$^{2}$ cm$^{-3}$, 
decreasing from right to left.}

\label{sette}
\end{figure*}

We have considered the sensitivity of the fine structure emission line 
intensities to H-nuclei densities in the range 10$^{2}$$\leq$ 
n$_{H}$$\leq$10$^{5}$ cm$^{-3}$. In Fig~\ref{sette} we show our model 
results for different values of density. The [C {\sc ii}]$_{158}$ /[O {\sc 
i}]$_{63}$ and [O {\sc i}]$_{63}$/[O {\sc i}]$_{145}$ ratios are both 
sensitive to density, however there is a higher sensitivity for the [C 
{\sc ii}]$_{158}$ /[O {\sc i}]$_{63}$ ratio. As expected, there is a 
substantial decline in the [O {\sc i}]$_{63}$/[O {\sc i}]$_{145}$ ratio as 
the density decreases from n$_{H}$$\sim$10$^{5}$ cm$^{-3}$ to 
n$_{H}$$\sim$9x10$^{3}$ cm$^{-3}$, and a steady decline in the [O {\sc 
i}]$_{63}$/[O {\sc i}]$_{145}$ ratio as the density decreases from 
n$_{H}$$\sim$9x10$^{3}$ while the model [C {\sc ii}]$_{158}$ /[O {\sc 
i}]$_{63}$ ratios increase initially with decreasing density until the 
density reaches a value of n$_{H}$$\sim$10$^{3}$ cm$^{-3}$ and then
decrease as the density decreases further. Clearly, the changes in 
the [O {\sc i}]$_{63}$/[O {\sc i}]$_{145}$ ratio at 10$^{5}$ cm$^{-3}$ 
are partly due to the critical density of the two lines (see 
Table~\ref{ncr}). Similarly, the decrease in the [C {\sc ii}]$_{158}$ /[O 
{\sc i}]$_{63}$ ratio as the density decreases is partially due to the 
density falling below the critical density of the [C {\sc ii}]$_{158}$
line.

\subsubsection{Cosmic ray ionization rate}

\begin{figure*}
\psfig{file=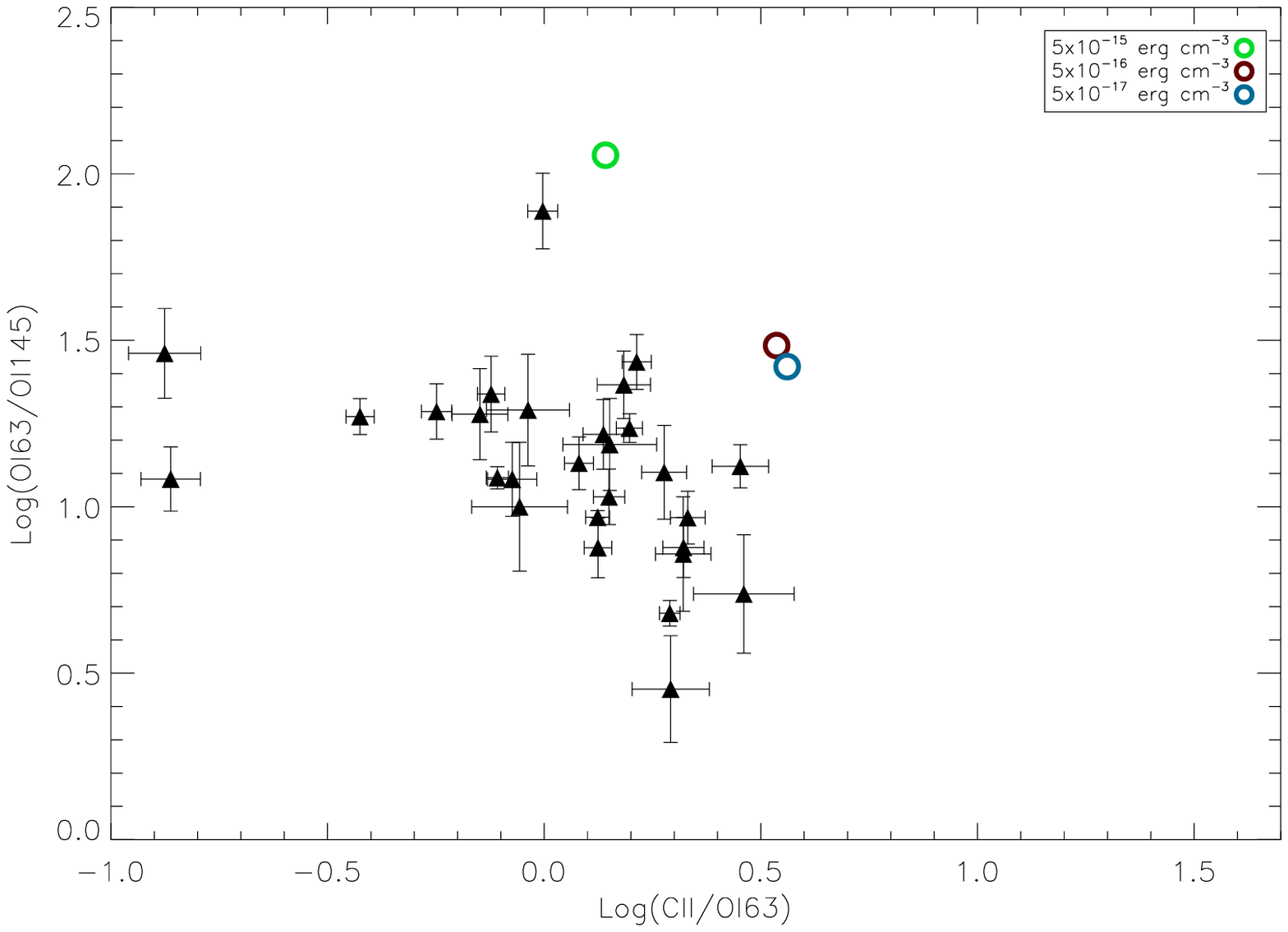,height=7cm,width=8.81cm}
\caption{Observed [C {\sc ii}]$_{158}$ /[O {\sc i}]$_{63}$ versus [O {\sc 
i}]$_{63}$/[O {\sc i}]$_{145}$ ratios, with uncertainties,
compared to models with differing cosmic ray ionization rates. The 
coloured circles represents models with cosmic ray ionization rates of 
$\zeta$=5x10$^{-15}$s$^{-1}$ (green circle),$\zeta$=5x10$^{-16}$s$^{-1}$ 
(claret circle) and $\zeta$=5x10$^{-17}$s$^{-1}$ (blue circle).}
\label{otto}
\end{figure*}

The cosmic ray flux is known to vary by over an order of magnitude in the 
Milky Way \citep{Schilke93}. Magnetic field lines can channel cosmic rays 
away from dense molecular cores; alternatively, the flux of particles in 
star forming regions can be many times higher than the canonical rate 
\citep{Schilke93}. The UCL\_PDR code does not have a separate 
treatment for X-ray Dominated Region
(XDR) effects that may be important in some galaxies 
\citep[see][]{Meij06}. However to a first approximation one may use an 
enhanced cosmic ray ionization rate to mimic the effects of XDRs 
(Bell $\&$ Hartquist et al. 2006). In Fig~\ref{otto} we show our model results for cosmic 
ray ionization rates in the range 
5x10$^{-15}$$\leq$$\zeta$$\leq$5x10$^{-17}$s$^{-1}$. The higher end 
of our range can represent a higher ionization rate \citep[see 
e.g.][]{McCall} or the effect of additional ionization due to X-rays. 
There is a substantial decline in the 
[O~{\sc i}]$_{63}$/[O~{\sc i}]$_{145}$ ratio as the cosmic ray ionization 
rate decreases from $\zeta$=5x10$^{-15}$s$^{-1}$ (green circle) to 
$\zeta$=5x10$^{-16}$s$^{-1}$ (claret circle) and it then decreases 
slightly for $\zeta$=5x10$^{-17}$s$^{-1}$ (blue circle);
the [C {\sc ii}]$_{158}$ /[O {\sc i}]$_{63}$ ratio decreases 
significantly from 
$\zeta$=5x10$^{-15}$s$^{-1}$ (green circle) to 
$\zeta$=5x10$^{-16}$s$^{-1}$ (claret circle) and then remains constant. 
In fact both the [C~{\sc ii}] and [O~{\sc i}]$_{63}$ fluxes decrease 
with a decrease in ionization rate, with [O~{\sc i}]$_{63}$ decreasing 
at a faster rate.

\section{Using [N~{\sc ii}] fluxes to estimate the ionized 
gas contribution to [C~{\sc ii}] fluxes}

Several studies have been made that show that significant [C~{\sc ii}] 
emission can arise from H~{\sc ii} regions 
\citep{Heiles94,Abel05,Kaufman06,Abel06}. Since the ionization potential 
from C$^{+}$ to C$^{2+}$ is 24~eV, the [C~{\sc ii}] 158~{$\mu$}m 
$^{2}$P$_{3/2}$ to $^{2}$P$_{1/2}$ line can be formed over a substantial 
part of a H~{\sc ii} region and according to \citet{Aannestad03} 
may contribute up to 1/3 of the total intensity in the line.
 
More recent work has indicated that approximately 25$\%$ of the observed 
[C~{\sc ii}] 158~$\mu$m emission may come from ionized regions 
\citep{Abel06}. This is particularly true where low-density H~{\sc ii} 
regions are adjacent to PDRs 
\citep{Heiles94,Abel05,Kaufman06,Abel07}. This effect can hamper the use 
of [C~{\sc ii}] emission as a pure PDR diagnostic in cases where ionized 
and PDR emission are observed in a single spectrum. Therefore, the 
contribution of [C {\sc ii}] emission from the ionized gas must be 
estimated. Such an estimate requires a separate model of the H~{\sc ii} 
region, although in recent years computional methods exist that allow the 
H~{\sc ii} region and PDR spectrum to be calculated self-consistentely 
\citep{Abel05,Kaufman06}. A study of S125 by \citet{Aannestad03} found 
that $\sim$40$\%$ of the [C~{\sc ii}] and $\sim$20$\%$ of the [O~{\sc i}] 
63-$\mu$m line intensities come from the ionized regions. Therefore, even 
though [C~{\sc ii}] emission is widely observed and is usually optically 
thin, its dependence on the properties of the H~{\sc ii} region can 
diminish its use as a PDR diagnostic.
The possibility that some of the [C~{\sc ii}] flux may come from 
H~{\sc ii} regions is not explicitly taken into account by the UCL\_PDR 
models. 

While ionized carbon, because of its ionization potential, can be found in 
both neutral gas and ionized gas clouds, species such as ionized nitrogen,
N$^+$, requiring an ionization potential of 14.53 eV, can arise only from 
H~{\sc ii} regions. Along with the [C~{\sc ii}] 158$\mu$m emission line, 
the [N~{\sc ii}] 122-$\mu$m and 205-$\mu$m lines are the 
brightest emission lines contributing to the total far-infrared emission
from our Galaxy (Wright et al. 1991). The ground state $^{3}$P 
term of the N$^{+}$ ion is split into the three $^{3}$P$_{2,1,0}$ levels 
from which the 122~$\mu$m and 205~$\mu$m lines arise. Therefore if a 
predicted H~{\sc ii} region value for the
[C~{\sc ii}]$_{158}$/[N~{\sc ii}]$_{122}$ flux ratio is available, the 
observed [N~{\sc ii}] 122-$\mu$m flux could be used to indicate the amount 
of [C~{\sc ii}] 158-$\mu$m emission arising from H~{\sc ii} regions
that are in the telescope beam.
 
The COBE FIRAS Galactic Plane spectral measurements yielded an integrated 
value of [N~{\sc ii}]$_{122}$/[N~{\sc ii}]$_{205}$=1.5 \citep{Wri91}. The 
same [N~{\sc ii}]$_{122}$/[N~{\sc ii}]$_{205}$ flux ratio was measured by 
\citet{Obi06} for the Great Carina nebula, which has N$_{e}$$\sim$30 
cm$^{-3}$. They used photoionization models for the Carina nebula to 
estimate [C~{\sc ii}]$_{158}$/[N~{\sc ii}]$_{122}$ = 1.6 for the H~{\sc ii} 
region. Table~\ref{nitrogen} lists the [N~{\sc ii}] 122-$\mu$m emission 
line fluxes measured by us from the ISO LWS spectra of 12 out of the 46 
sources listed in Table~\ref{sources} that showed a detection, 
together with the 
relative [C~{\sc ii}]$_{158}$/[N~{\sc ii}]$_{122}$ ratios and the 
estimated percentage of [C~{\sc ii}]$_{158}$ coming from the H~{\sc ii} 
region. The 12 sources that have detections of both [C~{\sc ii}] 
158~$\mu$m and [N~{\sc ii}] 122~$\mu$m have mean and median [C~{\sc 
ii}]$_{158}$/[N~{\sc ii}]$_{122}$ flux ratios of 10.2 and 5.9 
respectively. A H~{\sc ii} region [C~{\sc ii}]$_{158}$/[N {\sc 
ii}]$_{122}$ ratio of 1.6 implies that H~{\sc ii} regions contribute only 
16$\%$ (mean case) and 27$\%$ (median case) of the overall [C~{\sc ii}] 
158$\mu$m flux that is observed.
 
\begin{figure*}
\psfig{file=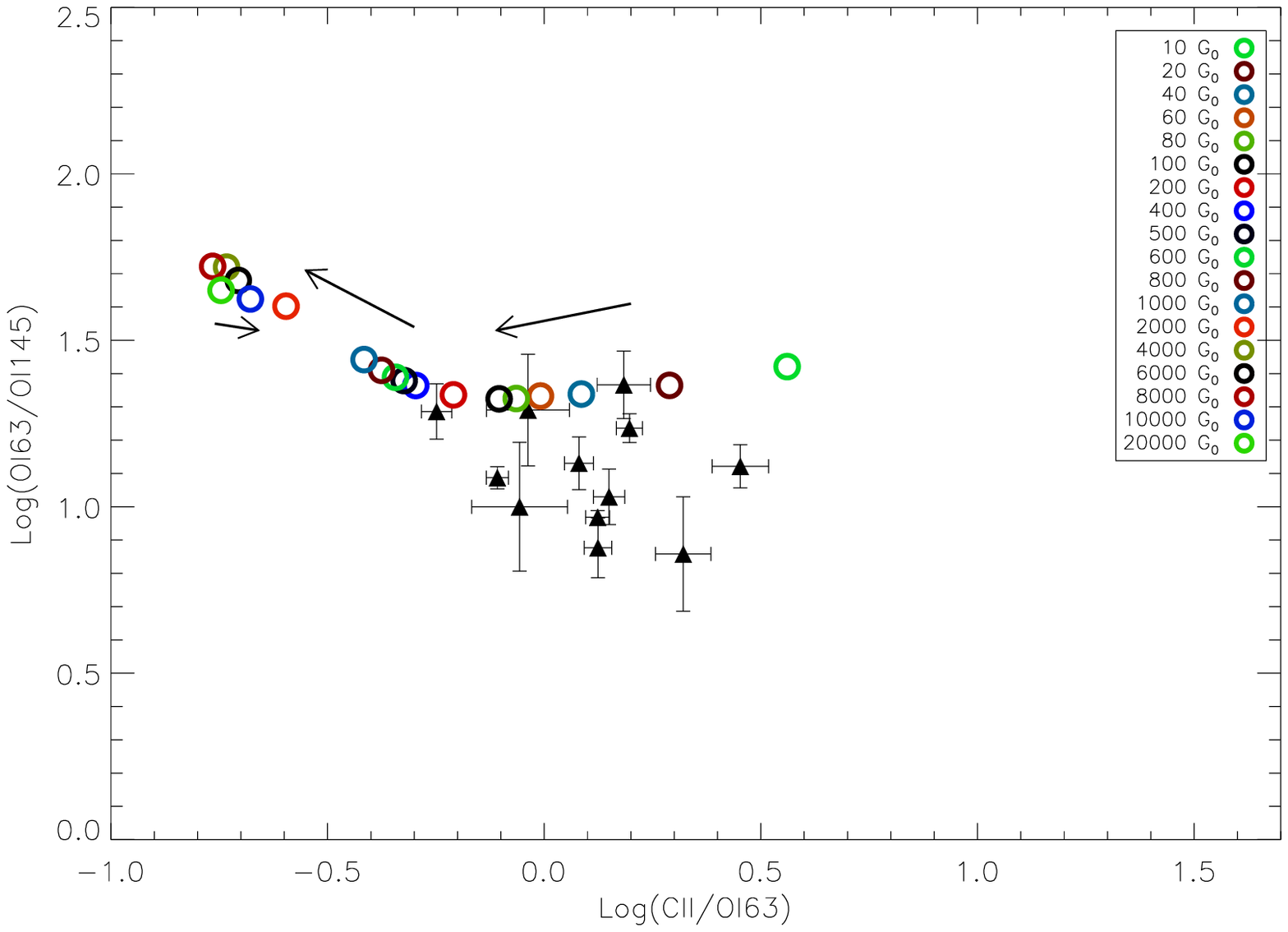,height=7cm,width=8.81cm}
\psfig{file=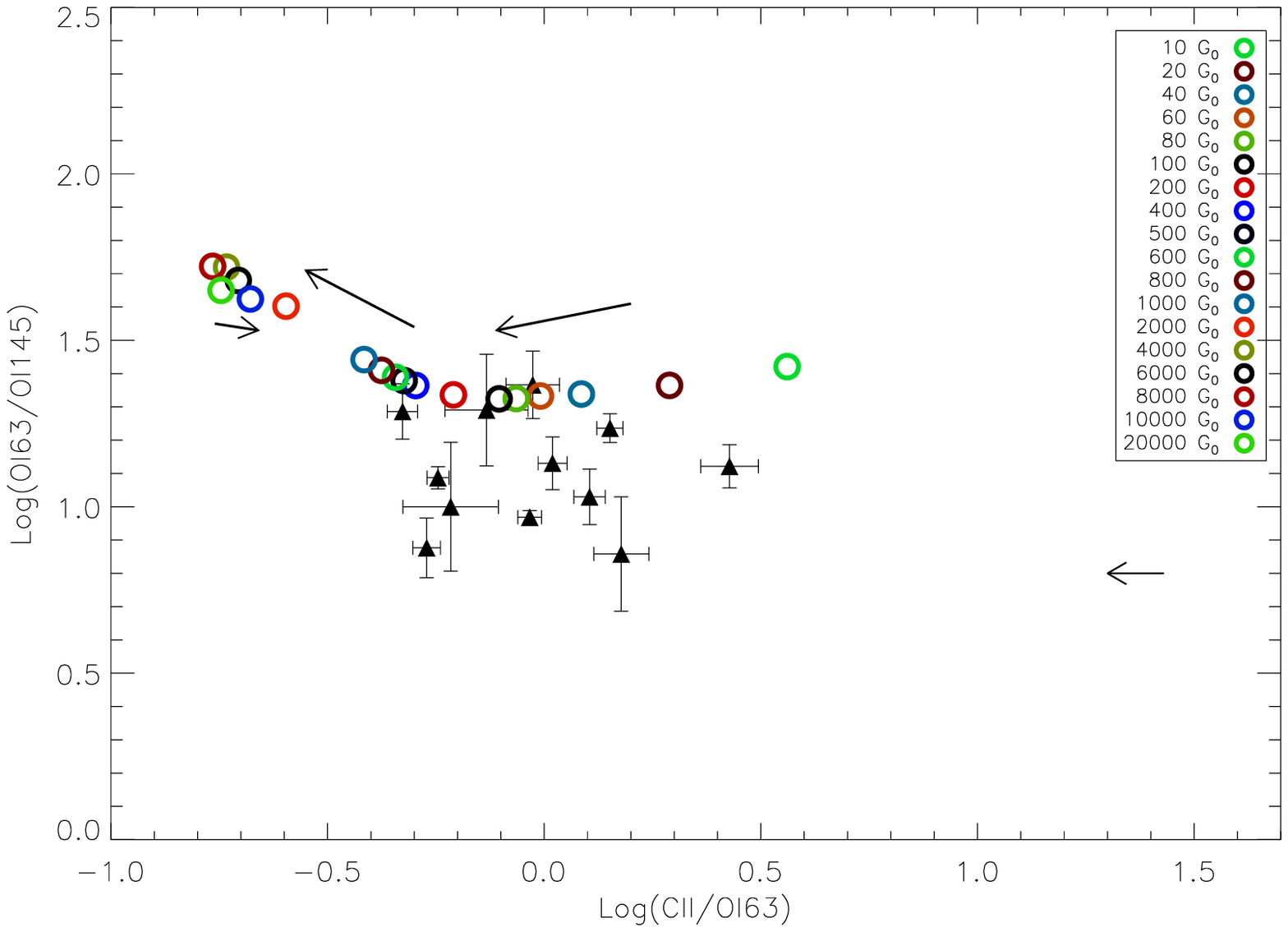,height=7cm,width=8.81cm}
\caption{Observed [C~{\sc ii}]$_{158}$ /[O~{\sc i}]$_{63}$ versus [O~{\sc 
i}]$_{63}$/[O~{\sc i}]$_{145}$ ratios, with uncertainties. In the left 
panel the observations do not take into account the likely percentage of 
[C~{\sc ii}] coming from H~{\sc ii} regions, while in the right panel the 
estimated contribution to the [C~{\sc ii}] flux is taken into account
for each source. The coloured points 
represent represent models with radiation field strengths in the range 
10$\leq$G$_{0}$$\leq$2x10$^{4}$ and the arrows indicate the direction of 
the increase of radiation field strength. The arrow on the right-hand side 
of the right panel corresponds to the median shift in the 
[C~{\sc ii}]$_{158}$/[N~{\sc ii}]$_{122}$ ratio, corresponding to 27$\%$ 
of the [C~{\sc ii}] emission originating from H~{\sc ii} regions (see 
Section 5).}
\label{nove}
\end{figure*}

We used the above predicted H~{\sc ii} region [C~{\sc ii}]$_{158}$/[N~{\sc 
ii}]$_{122}$ ratio of 1.6 along with the observed [N~{\sc ii}] 122-$\mu$m 
fluxes, to correct the observed [C {\sc ii}] 158$\mu$m flux of these 12 
sources for H~{\sc ii} region contributions. In Fig~\ref{nove} we overplot 
our model results, with the same physical parameters used in 
Fig~\ref{quattro}, on the observations. In the right-hand panel the 
observations have been corrected for the predicted H {\sc ii} region 
contribution to the 158-$\mu$m fluxes using the measured [N~{\sc ii}] 
122-$\mu$m fluxes. As 
shown in Fig~\ref{nove}, the model results appears to fit the corrected 
observations better, compared to Fig~\ref{quattro}, because the 
observations have been shifted to lower values of [C {\sc ii}]$_{158}$/[O 
{\sc i}]$_{63}$. The remaining discrepancy is attributed to [O~{\sc i}] 
63-$\mu$m self-absorption, as discussed in Section~6, below.

\begin{table*}
\caption{Line flux measurements, in units of 10$^{-14}$~W~m$^{-2}$, for 
extragalactic sources with ISO-LWS detections of the [N~{\sc 
ii}] 122-$\mu$m emission line, together with the observed [C~{\sc 
ii}]$_{158}$/[N~{\sc ii}]$_{122}$ flux ratio. Column 1 is the index number from 
Table~\ref{sources}. }
\label{nitrogen}
\begin{tabular}{llllll}
\hline
{\bf Index}& {\bf Source}& {\bf TDT No. of }& {\bf [N~{\sc ii}] 122$\mu$m}& {\bf \underline{[C~{\sc ii}]$_{158}$}} &{\bf [C~{\sc ii}]$_{158}$} \\
&          &       {\bf Observation }&  &    {\bf [N~{\sc ii}]$_{122}$} & {\bf from H~{\sc ii}}   \\
\hline
3& {\bf NGC 253}&  56901708& 1.72$\pm$0.18 & 2.7$\pm$0.4 & $\sim$60\%\\
\hline
5& {\bf Maffei 2}& 85800682& 0.188$\pm$0.014 & 5.7$\pm$0.7 & $\sim$28\% \\
\hline
6& {\bf NGC 1068}& 60500401& 0.356$\pm$0.026 & 5.3$\pm$0.6 & $\sim$30\% \\
\hline
10& {\bf NGC 1614}& 85501010& 0.021$\pm$0.004 & 9.7$\pm$2.5 & $\sim$17\% \\
\hline
11& {\bf NGC 2146}& 67900165& 0.152$\pm$0.006 & 16.3$\pm$6.4 & $\sim$10\% \\
\hline
13 & {\bf M 82}&  65800611& 2.20$\pm$0.17 & 5.9$\pm$0.6 & $\sim$27\% \\
\hline
18 & {\bf NGC 4039/9}& 25301107& 0.0468$\pm$0.0049& 8.08$\pm$1.09 & $\sim$20\% \\
\hline
26 & {\bf Cen A}&  63400464& 0.170$\pm$0.019  & 16.2$\pm$2.4 & $\sim$10\% \\
\hline
27 & {\bf NW Cen A}&  45400151& 0.097$\pm$0.011 & 28.8$\pm$5.7 & $\sim$5.5\% \\
\hline
28 & {\bf M 51}&  35100651& 0.228$\pm$0.027 & 4.2$\pm$0.7 & $\sim$38\% \\
\hline
29 & {\bf M 83}&  64200513& 0.231$\pm$0.066 & 5.2$\pm$1.7 & $\sim$31\% \\
\hline
30 & {\bf Circinus}&  10401133& 0.213$\pm$0.019 & 12.2$\pm$1.5 & $\sim$13\% \\
\hline
\end{tabular}
\end{table*}

\section{Fitting the observations}
\label{sec:results}

The clear discrepancy between the location of the observations and our 
models in Figures 3-8 can at least be partly attributed to self absorption 
in the [O~{\sc i}] 63-$\mu$m line. This is consistent with what 
\citet{Liseau06} found: namely that $\sim$65$\%$ of observed [O~{\sc 
i}]$_{63}$/[O {\sc i}]$_{145}$ emission line ratios from Milky Way PDRs 
are lower than can be explained by current models, which they attributed 
to optical depth effects in the [O~{\sc i}] 63-$\mu$m line. They partially 
attributed the low observed ratios to absorption by cold foreground 
O$^{0}$ in front of the 63-$\mu$m-emitting region, although other explanations 
such as very optically thick [O~{\sc i}] line emission could not be ruled 
out. 

The 
overall conclusion of \citet{Liseau06} was that because this ratio depends 
sensitively on models, [O~{\sc i}] emission has a limited diagnostic 
value. In fact most PDR models calculate the emergent flux from a 
plane-parallel slab of gas illuminated from one side. A galaxy has many 
PDRs at all orientations, and optical depth effects are non-negligible. In 
the approximation that the densest PDRs correspond to the shells of 
molecular clouds and that the [C~{\sc ii}] line emission and far-IR dust 
continuum emission are optically thin, while the [O~{\sc i}] 63-$\mu$m 
line is optically thick, then the [O~{\sc i}] 63-$\mu$m line will be seen only 
from 
the front side of each cloud while the [C~{\sc ii}] line arises from both 
the front and the near sides. The velocity dispersion from cloud to cloud, 
however, allows most [O~{\sc i}] 63-$\mu$m photons that have escaped their 
parent cloud to escape the galaxy entirely. This scenario implies that we 
should observe only a fraction of the [O~{\sc i}] flux and all of the 
[C~{\sc ii}] flux expected from PDR models \citep{Malhotra01}.

The observed [C~{\sc ii}]$_{158}$/[O~{\sc i}]$_{63}$ and [O~{\sc 
i}]$_{63}$/[O~{\sc i}]$_{145}$ ratios, in the selected range, do not 
appear to be traced properly by our UCL\_PDR model results although there 
is a 
common trend in the behaviour of these ratios, indicating that the 
observations are off-set compared to our model results. At first glance 
the model in the lower right panel of Fig~\ref{quattro}, where we varied the 
radiation field strength, appears to show the best fit to the 
observations. However, taking into account the effects of [O~{\sc i}] 
self-absorption and the [C~{\sc ii}] contribution from H~{\sc ii} regions, 
the 
models in the lower left panel of Fig~4 could also probably provide a fit 
to the 
observations. In fact both these effects would shift the models downwards 
and to the right relative to the observations. A similar trend could be 
found when we varied the density. Although at first glance the models in 
the top right panel of Fig~\ref{sette} do not appear to show a good fit to 
the observations, if one takes into account the effects of [O~{\sc 
i}]$_{63}$ self-absorption, which will shift observations 
downwards and to the left relative to the models, due to the decrease in 
the [O~{\sc i}] 63-{$\mu$}m flux, these models will probably provide a 
better fit to the observations. The models will be shifted downwards and 
to the right relative to the observations.  Unfortunately the results 
obtained varying the cosmic ray ionization rate and the metallicity 
suggest that the [C~{\sc ii}]$_{158}$/[O~{\sc i}]$_{63}$ versus [O~{\sc 
i}]$_{63}$/[O~{\sc i}]$_{145}$ ratio is not a good tracer for these 
parameters. We identified the best parameter values of our source sample 
as 10$^{4}$$\leq$n$_{H}$$\leq$9x10$^{4}$ cm$^{-3}$ and 
60$\leq$G$_{0}$$\leq$8x10$^{2}$, A$_{v}$= 10. These values were chosen 
while taking into account the effects of absorption in the [O~{\sc i}] 
63-{$\mu$}m line (See Section 6.1) and the contribution to [C~{\sc ii}] 
158-$\mu$m from H~{\sc ii} regions (Section 5).

\subsection{[O~{\sc i}] 63-$\mu$m self-absorption}

The intensity of the [O {\sc i}] 63{$\mu$}m line can be suppressed through 
self-absorption \citep[e.g.][]{Gonzo04}. The $^{3}$P$_{1}$ state, 
the upper level of the [O~{\sc 
i}] 63-{$\mu$}m line, is 228~K above the ground state and its critical 
density 
is n$_{crit}$ = 4.7x10$^{5}$ cm$^{-3}$ \citep{TH85}. Because of this, at 
typical interstellar cloud temperatures all oxygen atoms occupy the lowest 
level of their ground state, with J=2. This means that [O~{\sc i}] 
63-{$\mu$}m emission, originating from collisional excitation in a warm 
medium, could be absorbed by the abundant reservoir of ground state 
neutral oxygen that exists within the galaxy along the line of sight.  
The 63-$\mu$m line can also be absorbed by cold foreground material 
but this is less likely in the case of a face-on galaxy, where the 
neutral oxygen column density will be less along the line of sight than 
for an edge-on galaxy. For our 
sample of sources the majority are not edge-on 
galaxies, therefore we attribute most of the suppression of the 
[O~{\sc i}] 63-$\mu$m line to self-absorption in the PDRs in
which the emission is excited.

\subsubsection{The [O~{\sc i}] 63-$\mu$m line profile}

\begin{figure*}
\psfig{file=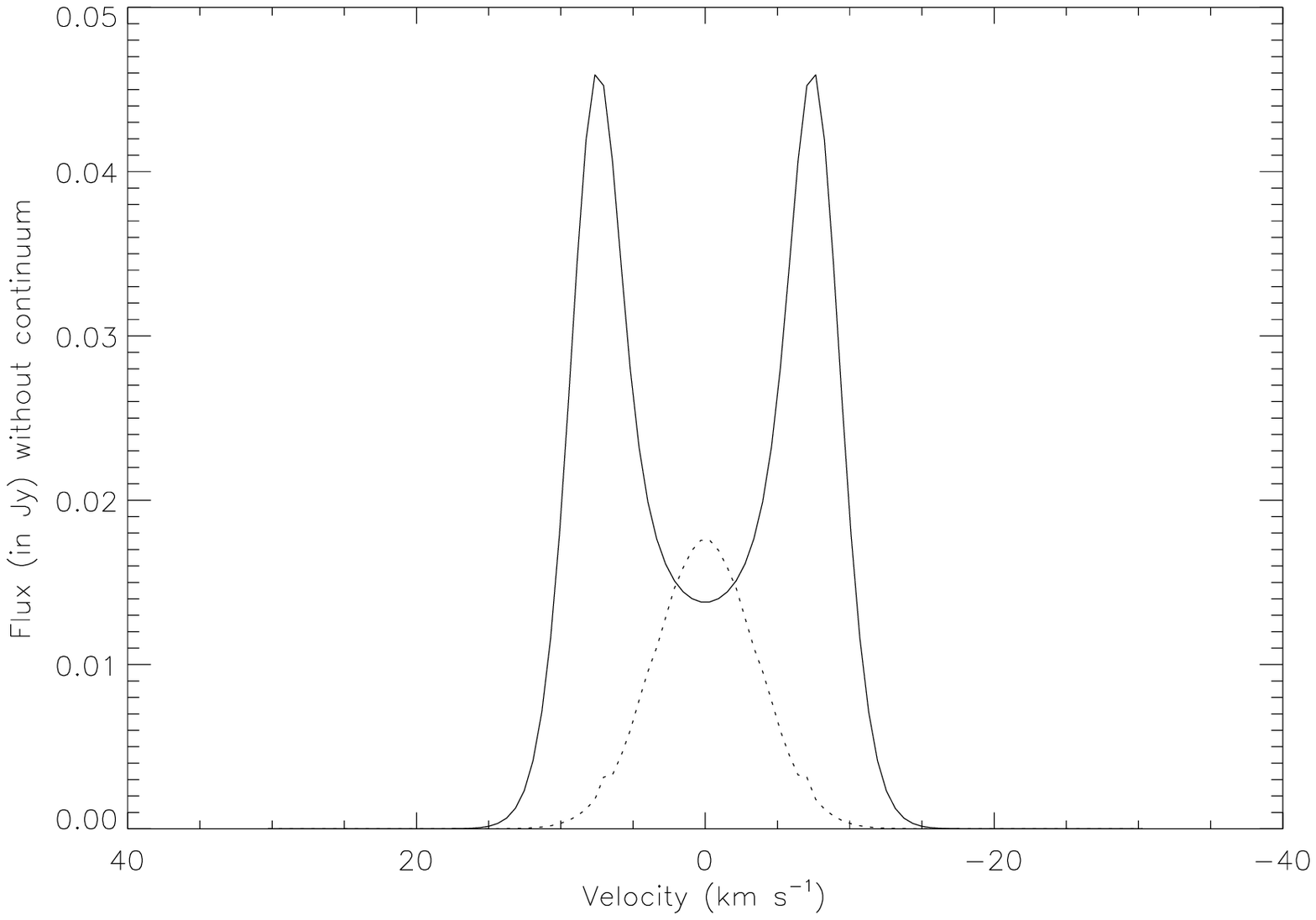,height=7cm,width=8.81cm}
\psfig{file=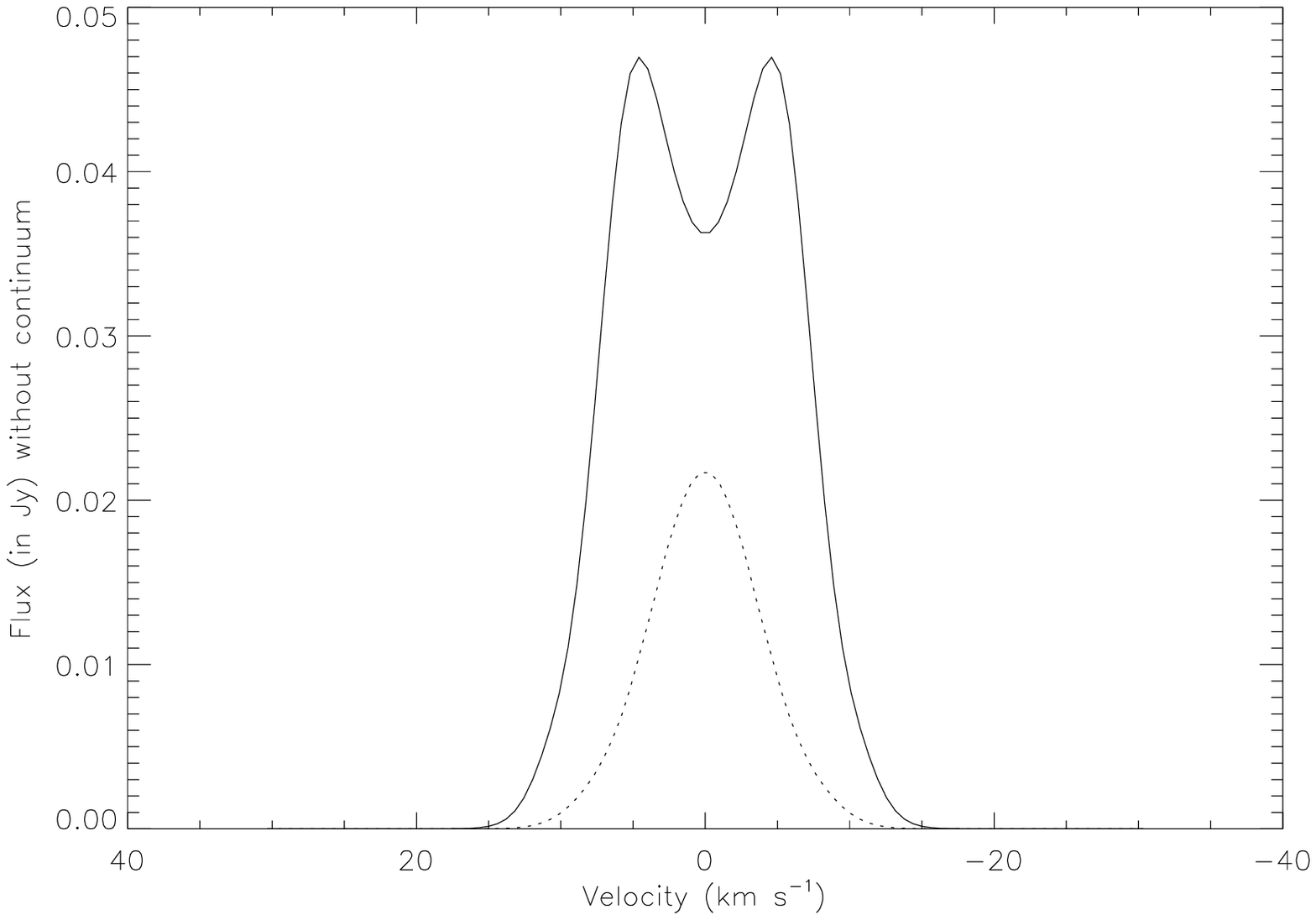,height=7cm,width=8.81cm}
\caption{Theoretical oxygen line profiles produced from a 20 pc GMC (left panel) and from a 5pc GMC (right panel), observed at a distance of 3.2 Mpc. The dotted line represents the [O {\sc i}]$_{145 \mu m}$ emission line profile and the solid line represents the [O {\sc i}]$_{63\mu m}$ emission line profile. The physical parameters of the UCL\_PDR model used for modeling the oxygen line profiles were $\zeta$=5x10$^{-17}$s$^{-1}$, 10$^{3}$ cm$^{-3}$, Z/Z$_{\odot}$=1, 200G$_{0}$ and t=10$^{7}$ yrs.}
\label{dieci}
\end{figure*}

In an attempt to investigate the effect of [O~{\sc i}] 63-$\mu$m 
self-absorption on the observations we have developed a new version of the 
Spherical Multi-Mol code (SMMOL, \citet{Rawlings01}). Originally SMMOL 
only considered a sphere of material illuminated by the standard ISRF
at the outer boundary of 
the sphere: we have now implemented the ability to 
insert a central illuminating source with an arbitrary SED. SMMOL uses an 
accelerated $\Lambda$-iteration (ALI) method to solve multilevel non-LTE 
radiative transfer problems of gas inflow and outflow. The code computes 
the total radiation field and the level populations self-consistently. At 
each radial point, SMMOL generates the level populations, the line source 
functions and the emergent spectrum from the cloud surface. This can then 
be convolved with the appropriate telescope beam.  A detailed description 
of the SMMOL radiative transfer model can be found in the appendix of 
\citet{Rawlings01}.  The coupling between the UCL\_PDR code and the 
radiation transfer code has been performed through an interface \footnote{See website: https://www.astro .uni-koeln.de /projects /schilke/ sites/www.astro.uni -koeln.de.projects.schilke/ files/ Viti\_Cologne09.pdf} that will  be presented in a forthcoming paper (Bayet et al. 2010, in prep).  We used the physical parameters from one of the UCL\_PDR models\footnote{Model result listed in Fig~\ref{quattro} (bottom left panel): $\zeta$=5x10$^{-17}$~s$^{-1}$, 10$^{3}$~cm$^{-3}$, Z/Z$_{\odot}$=1, 200G$_{0}$ and t=10$^{7}$~yrs.} and we chose an arbitrary distance of 
3.2~Mpc for the clouds modeled. In Fig~\ref{dieci} we show the SMMOL [O~{\sc i}] emission line profiles for a PDR with a diameter of 20~pc (left panel) and a PDR with a diameter of 5 pc (right panel).
In Fig~\ref{dieci} the effect of self-absorption is clearly seen in the [O~{\sc i}]$_{63\mu m}$ line profile (solid line) while there is no trace of self-absorption in the [O~{\sc i}]$_{145\mu m}$ line profile (dotted line).  The [O~{\sc i}]$_{63\mu m}$ emission line is 
still significantly brighter than the [O~{\sc i}]$_{145 \mu m}$ emission 
line, consistent with current observations. The line profile from a single 
20~pc GMC exhibits extremely strong [O~{\sc i}] 63-$\mu$m self-absorption. 
Without self-absorption in the [O~{\sc i}] 63-$\mu$m line, the 63/145 flux 
ratio would be 46.2. With self-absorption, the flux ratio is found to be 
8.5, a reduction of 0.73 dex. The effect is smaller for the 5~pc model; 
the 63/145 flux ratio is reduced from 12.6 to 7.9, i.e. by 0.20 dex. The 
log(63/145) $\sim$ 0.9 flux ratios predicted by both SMMOL models are in 
reasonable agreement with the observed ratios in e.g. Fig~\ref{nove}. 
Although our 20~pc and 5~pc diameter SMMOL models produce 
similar final 63- and 145-$\mu$m fluxes, their relative masses 
(1.0$\times10^5$ and
1.6$\times10^3$~M$_{\odot}$, respectively) would seem to make an ensemble
of the smaller clouds more plausible for matching the observed ratios.

Note that although all profiles in Fig~\ref{dieci} have been modeled with 
the same physical parameters, the difference between the two [O~{\sc 
i}]$_{63\mu m}$ emission line profiles is due to the difference between 
the inner and outer temperatures of the two clouds. The 20~pc GMC is 
approximately 100~K cooler than the 5~pc GMC at the centre. This is 
because the 20~pc GMC extends to A$_{v}$=19 mag, while the 5~pc GMC has a 
visual extinction of $\sim$5~mags. These differences account for the much 
stronger [O~{\sc i}]$_{63\mu m}$ self-absorption in the left panel. 
The [O~{\sc i}]$_{145\mu m}$ line, on the other hand, is approximately the 
same for both GMCs.

In order to qualitatively understand the behaviour of [O {\sc i}]$_{63 
\mu m}$ self-absorption in PDRs, we investigated its sensitivity to
variations in the density and radiation field strength. We find that an 
order of magnitude increase in the radiation field strength causes the 
inner and outer 
temperatures to be twice as high as those in the models plotted in Fig. 
10. This produces a higher flux and an increase in self-absorption in 
the  [O~{\sc i}]$_{63 \mu m}$ line.  Increasing the density by one order 
of 
magnitude causes an increase in the optical depth of the cloud, so that 
the degree of
[O~{\sc i}]$_{63 \mu m}$ self-absorption is larger than in the right-hand 
panel of Fig~\ref{dieci}, while the overall flux is reduced by an order 
of magnitude. We underline here that we are not attempting to model the 
actual line profiles that may be exhibited by these galaxies: in order to 
do so one would need to consider density gradients, multiple sources and 
the geometry of the galaxy. Our theoretical 
line profiles simply demonstrate the effect of the initial 
conditions on the degree of self-absorption in the [O~{\sc i}]$_{63 \mu 
m}$ line.

%

The [O~{\sc i}] 63/145 flux ratios predicted by our SMMOL models shift 
them downwards by 0.2-0.7~dex relative to the UCL\_PDR predictions,
bringing them into quite good agreement with the 
observed ratios plotted in Figs.~9 and 10. However, allowance for 
the 0.2-0.7~dex reduction in the [O~{\sc i}] 63-$\mu$m line strength
caused by self-absorption would also move the UCL\_PDR [C~{\sc 
ii}]/[O~{\sc i}]$_{63\mu m}$ model ratios in those figures to larger values.
We ran  SMMOL for the case of C$^+$ and we found, as expected, that 
the [C~{\sc ii}]$_{158\mu m}$ emission line profile did not show 
self-absorption in any of the models. 
In summary, we find that 20-80$\%$ of the intensity of the [O~{\sc 
i}]$_{63 \mu m}$ line can be suppressed through absorption, with 
the percentage depending strongly on the physical parameters of the PDR 
region. Comparing with the model trends shown in Fig.~9,
it would seem that a combination of high radiation field strengths 
and allowance for [O~{\sc i}] 63-$\mu$m self-absorption could provide the 
best match to the observed line ratios. [O~{\sc i}] line profile 
observations from the {\em Herschel Space Observatory} should allow a 
greater degree of discrimination amongst possible models. 
  
\section{Conclusions}
\label{sec:conclusions}

To fit the observed [C~{\sc ii}] 158~$\mu$m, [O {\sc i}] 63~$\mu$m and 
[O~{\sc i}] 145~$\mu$m emission line fluxes for 28 extragalactic sources, 
measured from archival ISO\--LWS spectra, we used a grid of 1702 PDR 
models from the UCL\_PDR code spanning a large range of densities, 
radiation field strengths, metallicities and cosmic ray ionization rates. 
We took into account the contribution to the observed [C~{\sc ii}] fluxes 
from H~{\sc ii} regions, using expected [C~{\sc ii}]$_{158}$/[N~{\sc 
ii}]$_{122}$ ratios for H~{\sc ii} regions to correct the [C~{\sc 
ii}]$_{158}$ fluxes of the 12 sources that had [N~{\sc ii}] 122~$\mu$m 
detections.

We found that the best fitting PDR models had 
10$^{4}$$\leq$n$_{H}$$\leq$9x10$^{4}$ cm$^{-3}$ and 
60$\leq$G$_{0}$$\leq$8x10$^{2}$. Consistent with the conclusions of 
\citet{Liseau06}, we found that the persistent discrepancy between the 
observed and modelled line flux ratios can at least partly be attributed 
to self-absorption in the [O~{\sc i}] 63-$\mu$m line. We used the SMMOL 
code to predict oxygen emission line profiles for several PDR models and 
found clearly self-absorbed [O~{\sc i}] 63-$\mu$m profiles, with 20-80$\%$ 
of the intensity suppressed, depending on the physical parameters of the 
PDR regions. A combination of high radiation field strengths and
[O~{\sc i}] self-absorption would appear to provide the best explanation
for the observed [C~{\sc ii}] 158~$\mu$m, [O~{\sc i}] 63~$\mu$m and 
[O~{\sc i}] 145~$\mu$m line flux ratios.

\section*{Acknowledgements}
We thank the referee, Dr N. Abel, for constructive comments
that helped improve the paper.
MV would like to thank Dr. Estelle Bayet for her help throughout this 
research. Samuel Farrens is thanked for comments. 

\bibliography{aamnem99,biblist.bib}

\bsp

\label{lastpage}

\end{document}